\documentclass[aip,jcp,reprint,groupedaddress]{revtex4-1}

\usepackage[english]{babel}
\usepackage{amsmath, amstext, amsfonts, amssymb}
\usepackage{dcolumn}
\usepackage{multirow}
\usepackage{bm}
\usepackage{graphicx}
\usepackage[dvipsnames]{xcolor}
\usepackage{color}
\usepackage{physics}
\usepackage{subdepth}
\usepackage[utf8]{inputenc}
\usepackage[T1]{fontenc}
\usepackage{hyperref}
\hypersetup{pdfencoding=unicode,
   colorlinks, 
   urlcolor =blue, 
   linkcolor=black, 
   filecolor=black, 
   citecolor=black 
  }
\inputencoding{utf8}
\usepackage[nolist]{acronym}
\usepackage[sort&compress]{natbib}

\newcommand{\supp}{Supplement:}%
\draft 

\begin{document}


%
%
\title{Multi-reference protocol for (auto)ionization spectra: application to molecules}


%
%
\author{Gilbert Grell}
\email[]{gilbert.grell@uni-rostock.de}
\affiliation{Institut f\"ur Physik, Universit\"at Rostock, Albert-Einstein-Str. 23-24, 18059 Rostock, Germany}
%
%
\author{Sergey I. Bokarev}
\affiliation{Institut f\"ur Physik, Universit\"at Rostock, Albert-Einstein-Str. 23-24, 18059 Rostock, Germany}

\date{\today}

\begin{abstract}
  We present the application of the spherically averaged continuum model to the evaluation of molecular photoelectron and resonant Auger electron spectra. In this model, the continuum wave function is obtained in a numerically efficient way by solving the radial Schr\"odinger equation with a spherically averaged molecular potential. Different approximations to the Auger transition matrix element and, in particular, the one-center approximation are thoroughly tested against experimental data for the CH$_4$, O$_2$, NO$_2$, and pyrimidine molecules. In general, this approach appears to estimate the shape of the photoelectron and autoionization spectra as well as the total Auger decay rates with reasonable accuracy, allowing for the interpretation of experimental results.%
\end{abstract}
%
%
\maketitle 
%

%
\begin{acronym}[RASSCF]
\acro{1CA}{One-Center Approximation}
\acro{2CA}{Two-Center Approximation}
\acro{ADC}{Algebraic Diagrammatic Construction}
\acro{AES}{Auger Electron Spectrum}
\acro{AIS}{Autoionization Spectum}
\acro{ANO}{Atomic Natural Orbital}
\acro{AS}{Active Space}
\acro{CAS}{Complete Active Space}
\acro{CASSCF}{\acl{CAS} \acl{SCF}}
\acro{CASPT2}{\acl{CAS} \acl{PT2}}
\acro{CI}{Configuration Interaction}
\acro{DFT}{Density Functional Theory}
\acro{DO}{Dyson Orbital}
\acro{ETMD}{Electron Transfer Mediated Decay} 
\acro{FANO-ADC}[Fano-ADC]{Fano-Stieltjes Algebraic Diagrammatic Construction}
\acro{FWHM}{Full Width at Half Maximum}
\acro{GTO}{Gaussian Type Orbital}
\acro{HF}{Hartree-Fock}
\acro{ICD}{Interatomic Coulombic Decay} 
\acro{IP}{Ionization Potential}
\acro{MCDF}{Multi-Configurational Dirac-Fock}
\acro{MCSCF}{Multi-configurational \ac{SCF}}
\acro{MO}{Molecular orbital}
\acro{QC}{Quantum Chemistry}
\acro{PES}{Photoelectron Spectrum}
\acro{PT2}{Second Order Perturbation Theory}
\acro{RAS}{Restricted Active Space}
\acro{RASSCF}{\acl{RAS} \acl{SCF}}
\acro{SCF}{Self-Consistent Field}
\acro{RAES}{Resonant Auger Electron Spectrum}
\acro{RASPT2}{\acl{RAS} \acl{PT2}}
\acro{RASSI}{\acl{RAS} State Interaction}
\acro{RPES}{Resonant \acl{PES}}
\acro{SD}{Slater Determinant}
\acro{SCI}{Spherical Continuum for Ionization}
\acro{SE}{Schr\"odinger Equation}
\acro{SO}{Strong Orthogonality}
\acro{XAS}{X-ray Absorption Spectrum}
\end{acronym}

%
%
\section{Introduction}
\label{sec_intro}

Photoionization and (resonant) autoionization processes encode the system's electronic structure into the kinetic energy of the ejected electrons, producing a \ac{PES} or a \ac{RAES}, respectively.
A particular advantage of these electron-out spectroscopies is that they map bound to continuum states and thus are not subject to selection rule suppression and are more flexible than optical spectroscopy.~\cite{Stolow_CR_2004, Hofmann2013, Huefner_Book}
For instance, the combination of X-ray valence \ac{PES} and \ac{RAES} has been used to unravel specific solute-solvent interactions of transition metal complexes in solutions.~\cite{Yepes_JPCB_2014, Golnak_SR_2016, Bokarev_WIRCMS_2019}
In addition, these processes can probe, initiate, and couple to complex electronic and nuclear dynamics.~\cite{Wang_PRA_2018, Rudenko_N_2017, Unger_NC_2017, Stumpf_NC_2016, Slavicek_JPCL_2016} 
Moreover, they can cause biological radiation damage by a cascade of autoionization events.~\cite{Howell_IJRB_2008, Yokoya_IJRB_2017}

Often the resulting spectra are feature-rich and difficult to interpret based on the experimental data alone.~\cite{Bokarev_WIRCMS_2019, Nisoli_CR_2017}
Hence, the development of theoretical methods for the simulation of \ac{PES} and \ac{RAES} has accompanied the experimental advancement during the last decades.
Notably, modelling ionization processes remains challenging until today, although the fundamental theory is been known for decades.~\cite{Aberg1982, Starace_CaRiMI_1982, Fano_PR_1961, Breit_PR_1954}
Here, the most general framework is the multi-channel scattering formalism,~\cite{Fano_PR_1961} treating continuum and bound states on the same level and including correlation effects between them.~\cite{Aberg_CaRiMI_1982, Burke2011}
At this level of accuracy, solving the bound and continuum problems in a B-spline basis,~\cite{Bachau_RPP_2001} specialized Gaussian basis sets,~\cite{Colle_PRA_1989, Colle_PRA_1993} or combination of both~\cite{Marante_JCTC_2017} has proven to be a versatile approach for atoms, diatomics, and first-row hydrides.~\cite{Martin_JPBAMOP_1999, Colle_PRA_1993, Colle_JPBAMOP_2004, Klinker_PRA_2018}
%
%
%

More approximate approaches are usually obtained in the single-channel scattering formalism.~\cite{Burke2011}
For example, the multi-centered B-spline static-exchange \ac{DFT} method~\cite{Toffoli_CP_2002, Catone_PRL_2012} has been applied to molecules reaching the size of small organometallic complexes.~\cite{Stener_TCA_2007, Catone_PRL_2012, Calegari_S_2014a, Plesiat_JPCA_2019}
%
Further, methods that project the molecular problem on a one-center expansion allow to evaluate \ac{PES} and \ac{RAES} of small molecular systems with remarkable accuracy.~\cite{Demekhin_OS_2007, Demekhin_PRA_2009, Banks_PCCP_2017}
A particular efficient scheme is obtained, if the molecular continuum orbital problem is approximated by an atomic one.~\cite{Siegbahn_CPL_1975, Larkins_AJP_1990, Fink_JESRP_1995, Travnikova_CPL_2009, Inhester_PRA_2016, Hao_SD_2015}
%
%
%
Here, the simplest approach is to model the outgoing electron as a free particle or a distorted wave corresponding to an effective Coulomb potential,~\cite{Oana_JCP_2009, Oana_JCP_2007, Grell_JCP_2015} which has been in particular applied to \ac{PES} studies of rather large molecules.~\cite{Mignolet_C_2013, Gunina_JPCA_2016, Mohle_JCTC_2018, Moguilevski_CPC_2017, Raheem_SD_2017}
Finally, an entirely different set of methods relies on an implicit continuum representation with Stieltjes imaging~\cite{Carravetta_JCP_2000, Gokhberg_JCP_2009, Kolorenc_JCP_2008, Stener_JoESaRP_1995} or a Green's operator formalism.~\cite{Schimmelpfennig_JESRP_1995}
%
%
However, energies of several hundreds of eV, targeted in X-ray \ac{PES} and \ac{RAES} necessitate large basis sets leading to demanding computations.

In all these methods, there is a certain trade-off between the accuracy that can be afforded for modelling the bound states and the continuum part on the one side and the size of the system on the other. 
On the one pole, the multi-channel methods are residing which are highly accurate but computationally expensive and, thus, allow for treatment of only small systems.
The other extreme is to neglect the multi-center molecular potential and use simplistic representations like the free-particle approach, sacrificing the accuracy in favor of computational feasibility.
%
%
%
Moreover, most of the methods described above have a single reference character and are not suited to treat systems possessing multi-configurational wave functions, e.g., open-shell and transition metal compounds or dynamics near conical intersections.
%
%
%


%

In this article, we employ a multi-reference \ac{QC} protocol and present the \ac{SCI} approach for the evaluation of photoionization cross sections and partial Auger decay rates for the case of molecules.
%
The central approximation in the protocol is that the angular structure of the molecular potential is averaged out, leading to spherically symmetric continuum orbitals that are obtained by numerically solving the radial Schr\"odinger equation.
Thus, it represents a compromise between the two extreme cases mentioned above.
This work is a logical continuation of our previous benchmark of the protocol for the atomic case of \ac{RAES} of the neon $1s^{-1}3p$ resonance, where it was shown to yield spectra and total decay rates in good agreement with experimental references.~\cite{Grell_PRA_2019}
Such an approach, being natural for atoms, requires a numerical justification for the non-spherically symmetric molecular case.
As is demonstrated here, the \ac{SCI} method indeed provides a valuable insight into the character of photoionization and autoionization molecular spectral features.
In addition, we investigate the performance of the popular \ac{1CA} to Auger decay,~\cite{Siegbahn_CPL_1975, Travnikova_CPL_2009, Inhester_PRA_2016, Holzmeier_JCP_2018} neglecting contributions from atoms other than the core-hole bearing one.

Concerning the applicability of the \ac{SCI} ansatz, the crucial role is expected to be played by the deviations of the molecular point symmetry from the spherical one.
In this respect, highly symmetric molecules should in general be more suitable objects for \ac{SCI} representation, whereas for highly non-spherical systems, e.g., linear or planar, \ac{SCI} model might represent a quite crude approximation.
Therefore, we start the discussion with the K-edge spectra of $\text{CH}_4$, being isoelectronic to neon studied previously,~\cite{Grell_PRA_2019} which is the simplest system out of the selected series as it contains only one heavy atom and possesses a high symmetry.
Increasing the complexity, we continue with the open shell molecules $\text{O}_2$ and $\text{NO}_2$, allowing for ionization channels corresponding to different spins of ionic remainder, and address the K-edges of oxygen and nitrogen, respectively. 
The largest object studied herein is the pyrimidine ($\text{C}_4\text{H}_4\text{N}_2$) which is an aromatic heterocyclic system, where we aim at description of nitrogen K-edge.
For O$_2$ and pyrimidine, there is ambiguity in selecting the origin of the spherically symmetric continuum orbital because of presence of two equivalent atoms, which makes them also interesting objects for testing.

{One should note that for such small molecules vibrational effects, showing up as, e.g. vibronic progressions in the (auto)ionization spectra, are of importance and alter the measured experimental spectrum.~\cite{Holzmeier_JCP_2018}
Furthermore, they can lead to emergence of a measurable signal, even if a purely electronic transition is forbidden, due to vibronic interactions~\cite{Kivimaki_JPBAMOP_1996} or show up in new features appearing due to the ultrafast dissociation.~\cite{Caldwell_JESRP_1994}
In this work, we considered only electronic effects since the main goal is to test our multi-reference protocol for the case of molecules.
The full treatment, however, would require the inclusion of the nuclear motion into the consideration.}

The article is organized as follows.
First, we briefly recapitulate the theory behind our approach in Section~\ref{sec_theory}, which is presented in detail elsewhere,~\cite{Grell_PRA_2019} and give the details of our computational setup in Section~\ref{sec_comp}.
The \ac{XAS}, \ac{RAES}, and valence \ac{PES} are discussed on a case to case basis for each molecule separately in Section~\ref{sec_res}.
The overall summary across the series of molecules under study is given in Conclusions, Section~\ref{sec_conclusion}.

\section{Theory}
\label{sec_theory}

In this section, we briefly present theory behind our method; the detailed description can be found in our recent article.~\cite{Grell_PRA_2019}
%
%
Atomic units (a.u.) are employed throughout  this article, if not stated otherwise.
%
%
%
%
We restrict ourselves to the limits of first order perturbation theory, dipole approximation, and length gauge for the absorption and photoionization cross sections $\sigma_{gi}$ and $\sigma_{g\alpha}$.
Further, the average over the molecular orientation with respect to the polarization vector of the incoming light is taken.
Finally, the nonradiative decay rates $\Gamma_{i\alpha}$ are estimated within the two-step model,~\cite{Wentzel_ZP_1927} neglecting the excitation process and the interference of decay and direct ionization pathways.
The resulting expressions for \ac{XAS}, \ac{PES}, and \ac{RAES} read
%
\begin{align}
\label{eq:sigAbs}
\sigma_{gi}       &=  \frac{4\pi^2}{3c}\omega \left|\mel{\Psi_i}{\boldsymbol{\mu}}{\Psi_g}\right|^2,\\
\label{eq:sigPES}
\sigma_{g\alpha}  &=  \frac{4\pi^2}{3c}\omega k\left|\mel{\Psi_\alpha}{\boldsymbol{\mu}}{\Psi_g}\right|^2,\\
\label{eq:partRate}
\Gamma_{i\alpha}  &=  2\pi \left|\mel{\Psi_\alpha}{\mathcal{H}-\mathcal{E}_i}{\Psi_i}\right|^2,
\end{align}
respectively.~\cite{Bransden1983, Starace_CaRiMI_1982, Aberg_CaRiMI_1982}
Here, $\ket{\Psi_g}$ denote the ground state, $\ket{\Psi_i}$ are intermediate core-excited states (resonances), and $\ket{\Psi_\alpha}$ are ionized continuum states (autoionization channels) of the system.
Further, $\boldsymbol{\mu}=-\sum_{u=1}^N\mathbf{r}_u$ is the $N$-electron dipole operator, $\mathcal{H}=\sum_u h_u+\sum_{u<v} 1/r_{uv}$ is the molecular Hamiltonian, containing one-electron $h_u$ and two-electron $1/r_{uv}$ parts respectively, and $\mathcal{E}_i$, $\omega$, and $k$ are the energies of $\ket{\Psi_i}$, of the incoming radiation as well as the wavenumber of the ionized electron, respectively.
%
%
%
%
To ensure that the total spin and its projection onto the quantization axis for the unionized system, $S$ and $M$, are conserved, the continuum states are constructed as
\begin{equation}
\label{eq:continuum}
\ket{\Psi_\alpha} = \sum_{M^+=-S^+}^{S^+}\sum_{\sigma=-\frac{1}{2},\frac{1}{2}} C_{S^+,M^+; \sigma}^{S,M}\ket{\Upsilon_{\alpha}^{\sigma M^+}}\,,
\end{equation}
where $\ket{\Upsilon_\alpha^{\sigma M^+}} = a^\dag_{\alpha,\sigma}\ket{\Psi_{f,M^+}^+}$ are channel functions corresponding to the bound cationic states $\ket{\Psi_{f,M^+}^+}$ with an additional electron, created by $a^\dag_{\alpha,\sigma}$, in the continuum orbital $\ket{\psi_{\alpha,\sigma}}$.
The $C_{S^+,M^+; \sigma}^{S,M}$ are the Clebsch-Gordan coefficients; note that $S^+$ and $M^+$ correspond to the total spin and its quantization axis projection of the ionic remainder.

The central approximation of the \ac{SCI} approach is to employ the spherically averaged molecular potential of the ion and thus spherically symmetric continuum orbitals
\begin{equation}
\label{eq:contOrb}
\psi_{\alpha,\sigma}(r,\Omega) = \frac{1}{r} w^{fk}_{l}(r)Y_{l}^{m}(\Omega)\zeta(\sigma),
\end{equation}
where spherical harmonics $Y_l^m(\Omega)$, $\zeta(\sigma)$, and $w^{fk}_{l}(r)$ correspond to the angular, spin, and radial parts of the wave function, respectively.
The compound channel index $\alpha=(f,l,m,k)$ contains the index of the ionic bound state $f$, the angular and magnetic quantum numbers $l$ and $m$, and the wave number $k=\sqrt{2\varepsilon_\alpha}$ of the outgoing electron.
For the evaluation of the spectra, one needs to sum over all decay channels $\alpha$ and in particular $l$ and $m$.
The energy of the outgoing electron is
\begin{equation}
\label{eq:energy}
\varepsilon_\alpha =
\left \{
\begin{array}{rl}
\omega + \mathcal{E}_g-\mathcal{E}_f, & \text{direct ionization}\\
\mathcal{E}_i-\mathcal{E}_f, & \text{Auger decay}\\
\end{array}
\right . .
\end{equation}
In contrast to the atomic case,~\cite{Grell_PRA_2019} for a molecule the origin of the continuum orbital, $\mathbf{r}_\text{c}$, and the molecular coordinate system in general differ.
In fact, the location of $\mathbf{r}_\text{c}$ is often ambiguous.
{Thus $\mathbf{r}_s = \mathbf{r}-\mathbf{r_\text{c}}$ defines the reference frame for Eq.~\eqref{eq:contOrb}.}
The functions $w^{fk}_{l}(r)$ are numerical solutions to the radial Schr\"odinger equation with the spherically averaged direct Coulomb potential $V_f^\text{J}(r) = V^\text{nuc}(r) + J_f(r)$ of the cationic state $\ket{\Psi_{f,M^+}^+}$ defined in spherical coordinates $\mathbf{r}_s=(r,\Omega)$ and containing nuclear $V^\text{nuc}(r)$ and electronic $J_f(r)$ parts.
They are computed using Numerov's method on a radial grid.
Asymptotically the solutions are constructed such as to fulfill the conditions
\begin{subequations}
	\begin{align}
	\label{eq:bcZero}
	w_l^{fk}(r\rightarrow 0) &= n r^{l+1}, \\
	\label{eq:bcInfty}
	w_l^{fk}(r\rightarrow \infty) &= \sqrt{\frac{2}{\pi k}} \Bigl( \cos{\delta^f_l(k)} F_l(\eta, kr ) \nonumber\\
	& \hspace{0.90cm}+ \left. \sin{\delta^f_l(k)} G_l(\eta, kr) \right) .
	\end{align}
\end{subequations}
Therein $\delta^f_l(k)$ are the scattering phases and $F_l$ and $G_l$ are the regular and irregular Coulomb functions,~\cite{Olver2010} respectively.
%

%
The nuclear part of $V_f^\text{J}(r)$, corresponding to the nuclear charges being smeared out over a sphere around the photoelectron origin, resembles the classical potential of charged hollow spheres with the radii $R_{\text{c}{A}} = |\mathbf{r}_\text{c} - \mathbf{R}_A|$ and charge $Z_A$
\begin{equation}
\label{eq:nucPot}
V^\text{nuc}(r) = \sum_A \left\{
\begin{array}{lr}
-\frac{Z_A}{R_{\text{c}A}}, & r < R_{\text{c}A} \\
-\frac{Z_A}{r},   & r \geq R_{\text{c}A}
\end{array}
\right. .
\end{equation}
The electronic part $J_f(r)$, however, is the electrostatic potential of the spherically averaged electron density of the respective ionized state, which has to be determined numerically within the $\mathbf{r}_s$ reference frame.~\cite{Grell_PRA_2019}
%
%
%

%
To keep this 
approach flexible with respect to the electronic structure method, we only require that the bound state wave functions are expressed as \ac{CI} expansions built on the \acp{MO} expressed in terms of conventional Gaussian type orbital basis sets.
Further, it is assumed that the neutral and cationic states have been obtained in separate calculations, comprising different relaxed sets of $N_\text{orb}$ spin orbitals $\{\varphi_i\}$ and $\{\varphi_i^+\}$, respectively.
Finally, the \ac{SO} approximation is employed throughout this work, since previous investigations~\cite{Grell_PRA_2019} have shown that quite reliable results can be obtained with it.
Hence, the overlap between the continuum and bound orbitals is neglected. 

For a particular ionization channel function this yields
\begin{equation}
\label{eq:PESME}
\mel{\Upsilon_{\alpha}^{\sigma M^+}}{\boldsymbol{\mu}}{\Psi_g } \stackrel{\text{SO}}{=}-\mel{\psi_{\alpha,\sigma}}{\mathbf{r}}{\Phi_{g\alpha}^{M^+}},
\end{equation}
and the Auger decay matrix element reads
\begin{align}
\label{eq:AugerME}
&\mel{\Upsilon_\alpha^{\sigma M^+}}{\mathcal{H}-\mathcal{E}_i}{\Psi_i} \\
\stackrel{\text{SO}}{=}& \mel{\psi_{\alpha,\sigma}}{h}{\Phi_{i\alpha}^{M^+}} + \underbrace{
	\sum_{q=1}^{N_\text{orb}} \mel{\psi_{\alpha,\sigma}\varphi^+_{q}}{\frac{1}{r_{12}}}{\Xi_{i\alpha}^{M^+,q}}
}_{r^{-1}\text{ coupling}} .\nonumber
\end{align}
%
%
Here, the matrix elements have been transformed to one and two-body integrals in terms of the corresponding Dyson orbitals $\ket{\Phi_{n\alpha}^{M^+}}$ ($n=i,g$) and two-electron reduced transition densities $\ket{\Xi_{i\alpha}^{M^+,q}}$.
%
Formal details and the numerical procedure to evaluate the continuum-bound matrix elements are elucidated in Ref.~\citenum{Grell_PRA_2019}.
%
A popular approximation to the full Auger matrix element ($\mathcal{H}$ coupling) is to disregard the one-electron terms, only accounting for the electronic Coulomb interaction ($r^{-1}$ coupling).
Further, the \ac{1CA} can be applied on top of the $r^{-1}$ coupling to reduce the computational demands.
%
Therein, the continuum orbital is placed on the core vacancy bearing atom and {non-local contributions from all other atoms} to the matrix element in Eq.~\ref{eq:AugerME} are neglected.
The performance of the $\mathcal{H}$ and $r^{-1}$ couplings as well as of the \ac{1CA} has been investigated here.
\section{Computational Details}
\label{sec_comp}
%
%
\begin{table*}[htb!]
	\caption{\label{tab:comp:QC}
		{\ac{QC} schemes used for the bound electronic structure calculations of the depicted molecules: employed basis set, active space, imaginary shift used in the respective \ac{PT2} calculation, and number of electronic states per charge/multiplicity of the molecule.
		The number of electrons in the active space corresponds to the neutral species.
    ANO-L~\cite{Widmark_TCA_1990} basis functions have been used in all cases.
    }
		%
		%
		%
		%
		%
		%
		%
	}
	\begin{ruledtabular}
		\begin{tabular}{l l l c l l}
			& Basis Set & Active Space & $\delta_\text{PT2}$ [a.u.] & Nr. of States & Comment \\
			\hline
			$\text{CH}_4$ & C: $[7s5p3d2f]$\footnotemark[1] & RAS$(10,1,1;1,4,19)$                & 0.01 & ${}^1\text{CH}_4$:   96  &
			SupSym\footnotemark[3]\\
			& H: $[3s2p1d]$   & RAS1: $1a_1$, RAS2: $2a_1$, $1t_2$  &      & ${}^2\text{CH}_4^+$: 308 & \\
			&                                 & RAS3: $3s(a_1)$, $3p(t_2)$, $3d(t_2)$, $3d(e)$, & & & \\
			&                                 &  \hspace{0.95cm} $4s(a_1)$, $4p(t_2)$, $4d(t_2)$, $4d(e)$, $5s(a_1)$ & & & \\
			\hline
			$\text{O}_2$  & $[5s4p2d1f]$  &  RAS$(16,2,0;2,8,0)$            & 0.1 & ${}^3\text{O}_2$:   826  &
			Linear\footnotemark[3] \\
			&                               &  RAS1: $1\sigma_u$, $1\sigma_g$ &     & ${}^2\text{O}_2^+$: 1008 & \\
			&                               &  RAS2: 2-3$\sigma_g$, 2-3$\sigma_u$, $1\pi_u$, $1\pi_g$ &     & ${}^4\text{O}_2^+$: 504  & \\
			\hline
			$\text{NO}_2$ & N: $[5s4p2d1f]$ & CAS$(19;11)$   & 0.3 & ${}^2\text{NO}_2$:   430 &  O(1s) frozen\\
			& O: $[5s4p2d1f]$ & CAS: 2-6$a_1$, $1a_2$, 1-2$b_1$, 2-4$b_2$  &     & ${}^1\text{NO}_2^+$: 825 & \\
			&                                 &              &     & ${}^3\text{NO}_2^+$: 990 & \\
			\hline
			C$_4$H$_4$N$_2$    & N: $[4s3p2d1f]$ & RAS$(34,1,1;2,15,8)$                        & 0.2 & ${}^1\text{C}_4\text{H}_4\text{N}_2$: 137    & SupSym\footnotemark[3]\\
			& C: $[4s3p2d1f]$ & RAS1: $\sigma(1a_1)$, $\sigma(1b_2)$                        &     &
			${}^2\text{C}_4\text{H}_4\text{N}_2^+$: 2297 &
			C(1s) frozen\\
			& H: $[3s2p1d]$   & RAS2: $\sigma(\text{5-10}a_1)$, $\sigma(\text{3-6}b_2)$, $\pi_1(1b_1)$, &     &   & \\
			&                                 & \hspace{0.95cm} $\pi_2(1a_2)$, $\pi_3(2b_1)$, $n_{\text{N}_-}(7b_2)$, $n_{\text{N}_+}(11a_1)$      &     &   & \\
			&                                 & RAS3: $\pi^*(2a_2)$, $\pi^*(\text{3-4}b_1)$,  &     &   & \\
			&                                 & \hspace{0.95cm} $\sigma^\text{Ryd}(\text{12-13}a_1)$, $\pi^\text{Ryd}(\text{5-6}b_1)$, $\pi^\text{Ryd}(3a_2)$   &     &   & \\
		\end{tabular}
		\footnotetext[1]{
			%
			ANO-L exponents~\cite{Widmark_TCA_1990} have been supplemented by $(8s6p6d4f)$ Rydberg exponents generated according to the procedure in Ref.~\citenum{Kaufmann_JPBAMOP_1989}.
			The $(22s15p10d7f)/[7s5p3d2f]$ contractions have been obtained with the \textsc{Genano} module~\cite{Almlof_JCP_1987} of \textsc{openMolcas}.~\cite{FernandezGalvan_JCTC_2019}, see \supp~Section~III.
			%
		}
		%
		%
		\footnotetext[3]{
			{Keyword used in the \acs{RASSCF} calculation in \textsc{MOLCAS 8.0} to prevent mixing of orbitals from different symmetries during the \ac{SCF} procedure.}
		}
		%
	\end{ruledtabular}
\end{table*}
%
%
%
%
%
For the molecules apart from oxygen, the geometries have been obtained with the \textsc{Gaussian09}~\cite{Frisch_Gaussian_2009} program at the B3LYP
/aug-cc-pVTZ level. 
For oxygen, we have employed the experimentally determined equilibruum distance of $r=1.208$~{\AA}.~\cite{Huber_MSaMS_1979}
The molecular symmetry has been restricted to the $T_d$ ($\text{CH}_4$) and $C_{2v}$ ($\text{NO}_2$ and pyrimidine) point groups. 
%

%
%
For the joint discussion of the \ac{XAS}, \ac{PES} and \ac{RAES} on equal footing, the underlying electronic structure calculation needs to satisfy all the demands for the involved electronic states.
In fact, this requires accurate predictions of the ground and core-excited states of the neutral species, as well as of the cationic valence-excited states.
For this purpose we have  selected the \ac{RASSCF}/\ac{RASPT2}~\cite{Malmqvist_JPC_1990, Malmqvist_JCP_2008} approach.
The active space is subdivided into the RAS1, RAS2, and RAS3 subspaces.
The RAS1 contains $o$ fully occupied orbitals, allowing for $h$ holes at maximum.
The occupation of the $c$ orbitals within RAS2 is not restricted, corresponding to a full \ac{CI} treatment within this subspace.
RAS3, however, contains $v$ virtual orbitals occupied by at most $p$ electrons.
Herein, we denote the total number of active electrons as $a$, which allows to uniquely specify each active space as RAS$(a,h,p;o,c,v)$.
If only the RAS2 space is used, it corresponds to the \ac{CASSCF}~\cite{Roos_CP_1980} calculation.
In this case, the active space is denoted as CAS$(a;c)$.
Due to the restricted configuration space central to these methods, the wave functions are often lacking dynamic correlation effects.
This is corrected to the second order of perturbation theory with the \acs{RASPT2}~\cite{Malmqvist_JCP_2008} and \acs{CASPT2}~\cite{Andersson_JPC_1990} methods.
%
%
%
All calculations have been conducted with a locally modified version of \textsc{MOLCAS}~8.0.~\cite{Aquilante_JCC_2016}
The \ac{QC} setups, i.e., the used basis sets, active spaces, number of states included in the state averaged \ac{SCF} procedure, special keywords, and imaginary shifts within the \ac{PT2} correction are detailed for each system in Table~\ref{tab:comp:QC}.
%

%
%
%
%
The Dyson orbitals and two-electron reduced transition densities in Eqs.~\eqref{eq:PESME} and \eqref{eq:AugerME} have been estimated using the biorthonormally transformed orbital and \ac{CI} coefficients~\cite{Malmqvist_IJQC_1986} for the neutral and cationic states.
\section{Results and Discussion}
\label{sec_res}
%
%

%

%
We have chosen to present all ionization spectra with respect to the electronic binding energies rather than the kinetic energies.
This introduces a common reference for \ac{PES} and \ac{RAES} irrespective of the photon energy at which they have been obtained, simplifying the analysis.
%
To allow for a uniform comparison, both the simulated results and the digitized experimental spectra have been aligned with respect to the  lowest vertical \acp{IP} of each molecule, taken from Refs.~\citenum{Bieri_JESRP_1979, Edqvist_PS_1970, Katsumata_CP_1982, Bolognesi_JCP_2012}.
In case of methane, the $2a_1^{-1}$ \ac{IP} has been used because the respective bands are narrower than the lowest $1t_2^{-1}$ ones.
%
Note that for the open shell systems  $\text{O}_2$ and $\text{NO}_2$ each ionization branch differing in the spin of the final ion was shifted individually in our simulated spectra.
The calculated \ac{XAS} have been shifted as well.
All applied shifts are summarized in Table~\ref{tab:res:shift}.
The broadening parameters for the \ac{XAS}, \ac{PES}, and \ac{RAES} of all molecules, tuned for the best agreement between our theory and the reference, are detailed in \supp\ Section II.
%
%
%
\begin{table}
  \caption{\label{tab:res:shift}
  The respective reference \acp{IP} and shifts (in eV) for the alignment of the theoretical and experimental ionization and absorption spectra.
  The experimental shifts have been applied to the indicated properties (see respective footnotes), whereas the theoretical ones have been uniformly applied to all \ac{PES} and \ac{RAES} for each system.
  }
\begin{ruledtabular}
  \begin{tabular}{c c c c  c}
    & \mbox{vertical \ac{IP}}\footnotemark[1]
    & $\Delta^{\text{Ion}}_\text{Theory}$ & $\Delta^{\text{Ion}}_\text{Exp}$  & $\Delta^\text{XAS}_\text{Theory}$\\
    \hline
    $\text{CH}_4$                    & 22.90~\cite{Bieri_JESRP_1979} ($2a_1^{-1}$)
    & -0.670   & 0.110\footnotemark[2]      & -0.850\\
    & &        & -0.260\footnotemark[3]~(B) & \\
    & &        & -0.300\footnotemark[3]~(D) & \\
    & &        & 0.200\footnotemark[3]~(H)  & \\
    & &        & 0.060\footnotemark[3]~(L)  & \\
    {$\text{O}_2$ }                    & 12.33~\cite{Edqvist_PS_1970} (${}^2\text{O}_2^+$)
    & -0.120   & -0.100\footnotemark[2]     &  0.294\\
                                     & 16.70~\cite{Edqvist_PS_1970} (${}^4\text{O}_2^+$)
    & -0.110   &        --\footnotemark[3]                    & \\
    $\text{NO}_2$                    & 11.25~\cite{Katsumata_CP_1982} (${}^1\text{NO}_2^+$)
    & -0.056   & 0.125\footnotemark[2]      & -0.938\\
                                     & 13.02~\cite{Katsumata_CP_1982} (${}^3\text{NO}_2^+$)
    & -0.245   & 0.100\footnotemark[3]      & \\
    $\text{C}_4\text{H}_4\text{N}_2$ & 9.81~\cite{Bolognesi_JCP_2012}
    & 0.200    &         --            &  2.680\\
  \end{tabular}
  \footnotetext[1]{The final state of the ion is given in parentheses.}
  \footnotetext[2]{\ac{PES}}
  \footnotetext[3]{\ac{RAES}}
\end{ruledtabular}
\end{table}

\subsection{Methane}
\label{sec_res:ch4}
%
\begin{figure*}
  \includegraphics{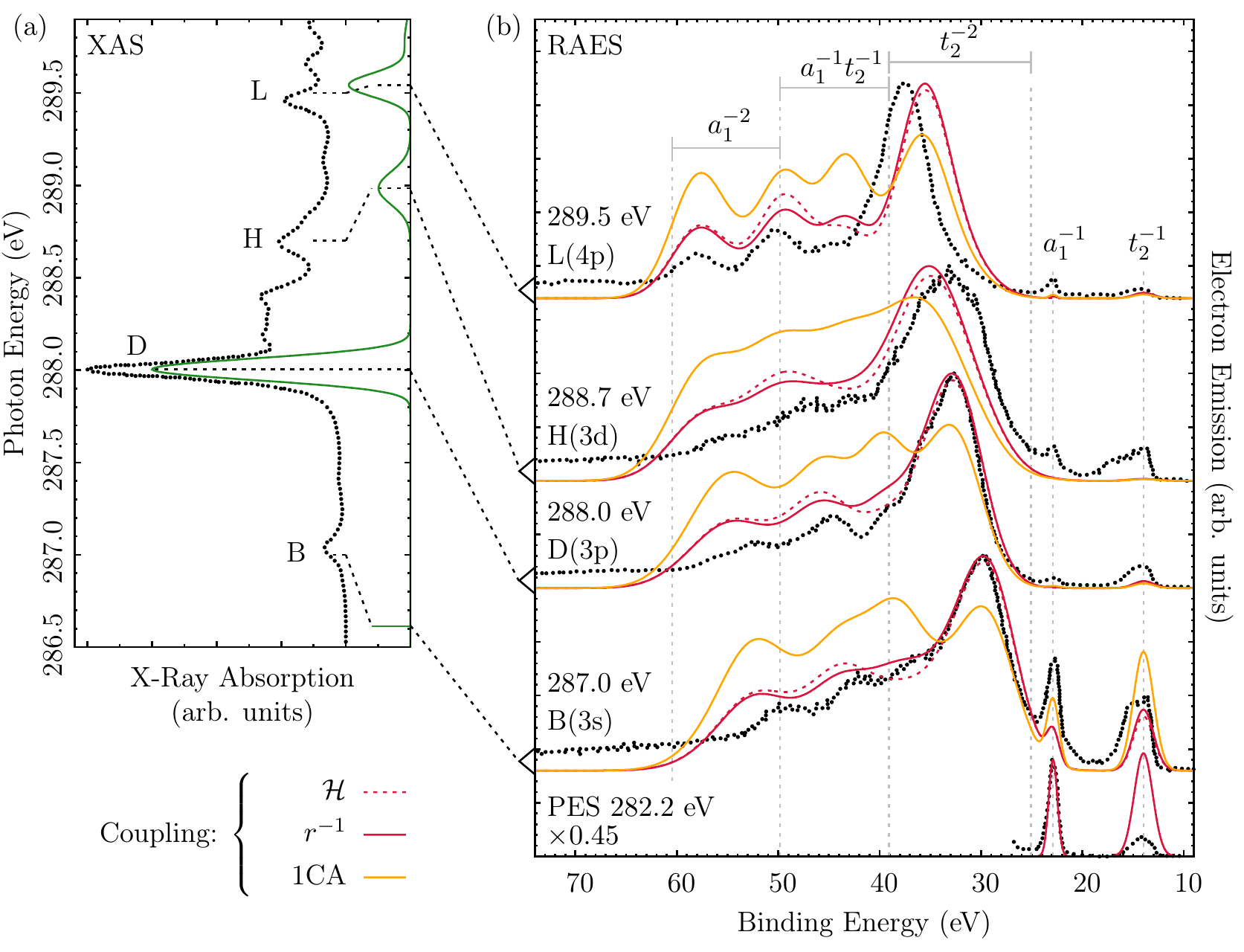}
  \caption{\label{fig:res:ch4}
    (a) Calculated \ac{XAS} of methane at the carbon K-edge (green, solid) compared to the experimental result (black, dotted).
    The position of the dipole-forbidden excitation to the $3s(a_1)$ orbital at $286.6$~eV is indicated by a single stick.
    The assignment of the experimental data follows Refs.~\citenum{Ueda_CPL_1995} and \citenum{Schirmer_PRA_1993} and has been connected to the corresponding core-excited states in our calculation (dashed).
    (b) \ac{RAES} obtained with our method using the \ac{SCI} model together with the respective measured spectra.
    The key for the coupling types is given under panel (a).
    All simulated spectra for one resonance have been normalized with the same constants to underline their differences.
    Dominant Rydberg contributions obtained in our calculation (in paranthesis) as well as the excitation energies used in the experiment are given for guidance.
    The experimental and theoretical valence \ac{PES} taken at a photon energy of $282.2$~eV are shown at the bottom as well.
    %
    %
    Note that all experimental data in this figure have been digitized from Kivimaeki et al.~\cite{Kivimaki_JPBAMOP_1996}
    Shifts and broadening parameters are detailed in Table~\ref{tab:res:shift} and the main text.
  }
\end{figure*}
$\text{CH}_4$ has been chosen because it is isoelectronic to the previously investigated neon atom~\cite{Grell_PRA_2019} and thorough studies regarding its \ac{XAS}, \ac{PES} and \ac{RAES} have been carried out both
experimentally~\cite{Kivimaki_JPBAMOP_1996, Ueda_CPL_1995, Schirmer_PRA_1993} and theoretically.~\cite{Higashi_CP_1982, Kvalheim_CPL_1982}
In addition, its tetrahedral symmetry with the central carbon atom closely resembles spherical symmetry, indicating that the \ac{SCI} approach could be well suited to study ionization processes in this case.
In Fig.~\ref{fig:res:ch4}, we compare our theoretical  results for the \ac{XAS}, \ac{PES}, and \ac{RAES} from various resonances to experimental reference data digitized from Kivim\"aki et al.~\cite{Kivimaki_JPBAMOP_1996}
Therein, the ionization spectra have been recorded using a magic angle geometry,~\cite{Feldhaus_RSI_1992} eliminating angular anisotropy effects that have not been covered in our angle-integrated spectra.
%

%
\paragraph{\ac{XAS}}
\label{sec_res:ch4:xas}
%
Fig. 1(a) contains the experimental~\cite{Kivimaki_JPBAMOP_1996} and calculated \ac{XAS}.
Both spectra are normalized to the D$(3p)$ peak and the experimental data are shifted by $2$ units of intensity for clarity.
Below we use the notation of bands from Ref.~\citenum{Kivimaki_JPBAMOP_1996}.
Note that we always omit the core orbital from the labeling of resonances here and in the following, unless explicitly required.
%
%
%
%

The experimental spectrum  comprises a wealth of features.
The assignment of them is complicated by the strong vibronic coupling especially for higher transition energies,~\cite{Schirmer_PRA_1993, Ueda_CPL_1995}  which is not included in our consideration.
%
%
%
%
For instance, the B$(3s)$ peak is due to the dipole forbidden $3s(a_1)$ core excitation which becomes allowed due to a coupling to $t_2$ vibrational modes with a frequency of $0.374$~eV.~\cite{Schirmer_PRA_1993}
This fits well to the offset of -0.4~eV in the resonance position predicted in our calculation.
The remaining peaks have been assigned as follows.
D$(3p)$ corresponds to the $3p(t_2)$ resonance 
and according to the literature~\cite{Ueda_CPL_1995, Schirmer_PRA_1993} should be free of vibrational effects. 
H$(3d)$ is a $3d(t_2)$ resonance in accordance with Ref.~\citenum{Ueda_CPL_1995} and is shifted by $0.3$~eV with respect to the experiment.
However, it may also be ascribed to a mixed $3d(t_2)$/vibrationally excited $3p(t_2)$ character which might explain the obtained shift.~\cite{Kivimaki_JPBAMOP_1996}
%
%
Finally, L$(4p)$ calculated with a slight offset of $0.1$~eV is assigned to the $4p(t_2)$ resonance.~\cite{Ueda_CPL_1995, Schirmer_PRA_1993} 
However, because of the aforementioned strong vibronic coupling other electronic states might contribute to this peak as well.~\cite{Schirmer_PRA_1993}
%
Generally, taking into account the absence of vibrational effects in our calculations, the relative energetic positions of the Rydberg resonances and their \ac{XAS} intensities are in good agreement with the experiments.
%
%
\paragraph{\ac{PES} and \ac{RAES}}
\label{sec_res:ch4:raes}
%
Panel (b) of Fig.~\ref{fig:res:ch4} shows the theoretical and experimental~\cite{Kivimaki_JPBAMOP_1996} \ac{RAES} taken at the indicated resonance energies. 
Further, the 
valence \ac{PES}, recorded 
at $282.20$~eV are depicted.
%
%
Complete, $\mathcal{H}$, and approximate, $r^{-1}$, couplings as well as the \ac{1CA} of the $r^{-1}$ coupling were employed to evaluate the partial decay rates.
To ensure convergence of the spectral intensities with respect to the continuum orbital angular momentum $l$, we took into account partial waves up to $l_\text{max}=5$ (\ac{1CA} \ac{RAES}) and $l_\text{max}=12$ (\ac{PES}, $\mathcal{H}$, and $r^{-1}$ \ac{RAES}.
This is natural, since the \ac{1CA} results correspond to matrix elements centered at one atom and thus converge faster in $l$.
%
%

%
%
%
%
Finally, the \ac{PES} and \ac{RAES} have been normalized to the $a_1^{-1}$ and $t_2^{-2}$ features, respectively.
Note that the normalization constants determined for the $r^{-1}$ \ac{RAES} have been used for the \ac{1CA} and $\mathcal{H}$ spectra as well to underline their relative differences.
The valence \ac{PES} contains two peaks corresponding to electron emission from the $2a_1$ and $1t_2$ valence orbitals, leading to single hole states depicted as $a_1^{-1}$ and $t_2^{-1}$.
Apparently, our model can not reproduce the relative \ac{PES} intensities of these peaks correctly at this excitation energy and overestimates the intensity of the $t_2^{-1}$ feature by a factor of $4$.
A study of the \ac{PES} with different model potentials (\supp~Fig.~S1) has revealed that the spectrum depicted in Fig.~\ref{fig:res:ch4}, employing $V_f^\text{J}(r)$, yields in fact the best agreement  possible  within our present model. 
%
A further improvement would require a non-spherical continuum model that accounts for the true molecular symmetry.
However, the $a_1^{-1}$ : $t_2^{-1}$ intensity ratio is quite sensitive to the photon energy,~\cite{Banna_CPL_1975, Backx_JPBAMP_1975} which suggests that this is a somewhat difficult case.
The \ac{RAES} can roughly be divided into the participator and spectator decay regions, below and above $25$~eV binding energy.
Participator decay involves the excited electron in the decay process, leading to single hole states that are the main \ac{PES} features as well.
Here, the corresponding states are $a_1^{-1}$ and $t_2^{-1}$, indicated by the dashed lines.
Spectator decay, in contrast, leaves the excited electron intact and leads to states with two holes in the valence shell.
Similar states result from normal Auger decay as well, albeit the additional electron introduces energetic shifts in the resonant case.
The dominant character of the respective two hole states obtained in our calculation is indicated at the top of panel (b).
Clearly the spectator decay is dominating the \ac{RAES}.
The main peak located in the range $25-40$~eV is due to $t_2^{-2}$ target states for all resonances.
The high energy tail however generally comprises three features for all but the H resonance, where vibrational effects lead to a stronger broadening that hides the detailed structure in the experimental data.
With decreasing binding energy, the first feature may be assigned to $a_1^{-2}$ states, while the latter two can be assigned to states of $a_1^{-1}t_2^{-1}$ character. 
The participator region is most prominent for the B$(3s)$ resonance.
Fig.~\ref{fig:res:ch4} (b) demonstrates that the theoretical spectra obtained with $\mathcal{H}$ and $r^{-1}$ coupling reproduce the experimental \ac{RAES} for all resonances with good accuracy.
%
Note that using a simple effective Coulomb potential $-1/r$ to obtain the continuum orbitals leads to considerably worse agreement, see~\supp~Fig.~S2.
Further, employing an additional radial Slater exchange term into the potential does not improve the results, see~\supp~Fig.~S3.
%
With respect to the $r^{-1}$ results, the $\mathcal{H}$ coupling introduces just a slight redistribution of intensity from the high energy flank of the $t_2^{-2}$ peak to the center of the high energy tail {(about 45-50\,eV)}.
In contrast, employing the \ac{1CA} leads to an overestimation of the tail {(>40\,eV)} and participator regions, whereas the main feature is underestimated.
Thus, it seems that non-local contributions from the hydrogen atoms can not be disregarded and the \ac{1CA} is not a suitable approximation when evaluating \ac{RAES} of methane with the \ac{SCI} approach.
In the remainder of this section, we will only refer to the results obtained with $\mathcal{H}$ and $r^{-1}$ coupling.
While the overall agreement for the $\mathcal{H}$ and $r^{-1}$ couplings is very good, the following differences remain.
For all resonances, the $a_1^{-2}$ and  $a_1^{-1}t_2^{-1}$ regions are slightly overestimated with respect to the main $t_2^{-2}$ feature.
%
Further, the features in the tail region are {blue shifted} by roughly $3.0$~eV and $2.0$~eV for the B$(3s)$ and D$(3p)$ resonances, while for the H$(3d)$ resonance the whole spectrum appears to be {blue shifted} by $1.0$~eV.
In contrast for L$(4p)$, the $t_2^{-2}$ feature appears {red shifted} by $2.0$~eV, which decreases towards higher binding energies to $0.5$~eV for the $a_1^{-2}$ peak.
We attribute these shifts to the fact that our active space does not allow to include enough electron correlation to represent the highly excited states of $\text{CH}_4^+$ with the same accuracy as the lower ones.
In addition, the lack of nuclear effects prohibits the description of shifts due to vibronic coupling in the resonances as well as the ionized states.
Finally, the $a_1^{-1}$ and $t_2^{-1}$ participator peaks are barely present for the H$(3d)$ and D$(3p)$ resonances and the $a_1^{-1}$ peak seems to be generally underestimated with respect to the $t_2^{-1}$ one.
This discrepancy is most probably due to the two-step approach to resonant Auger emission employed herein, disregarding the excitation process and the interference between photoionization and the Auger decay terms.
Further, the deficiencies of our \ac{SCI} model to describe the correct $a_1^{-1}$ : $t_2^{-1}$ intensity ratio in the \ac{PES}, might translate to the participator decay as well.
We conclude this discussion  with the observation that the relative energetic positions and intensities of the features in the methane B$(3s)$, D$(3p)$, H$(3d)$, and L$(4p)$ \ac{RAES} can be reproduced quite well with the \ac{SCI} approach, using the $V_f^\text{J}(r)$ potential together with $\mathcal{H}$ or $r^{-1}$ coupling, while vibrational effects can be disregarded in a first approximation.
The \ac{1CA}, however, is not enough to reproduce the relative intensities.
Further, an improved model of the continuum orbital seems to be required to recover the  $a_1^{-1}$ : $t_2^{-1}$ intensity ratio in the \ac{PES} at this energy.
%
%
%
\subsection{Molecular oxygen}
\label{sec_res:o2}
\begin{figure*}
  \includegraphics{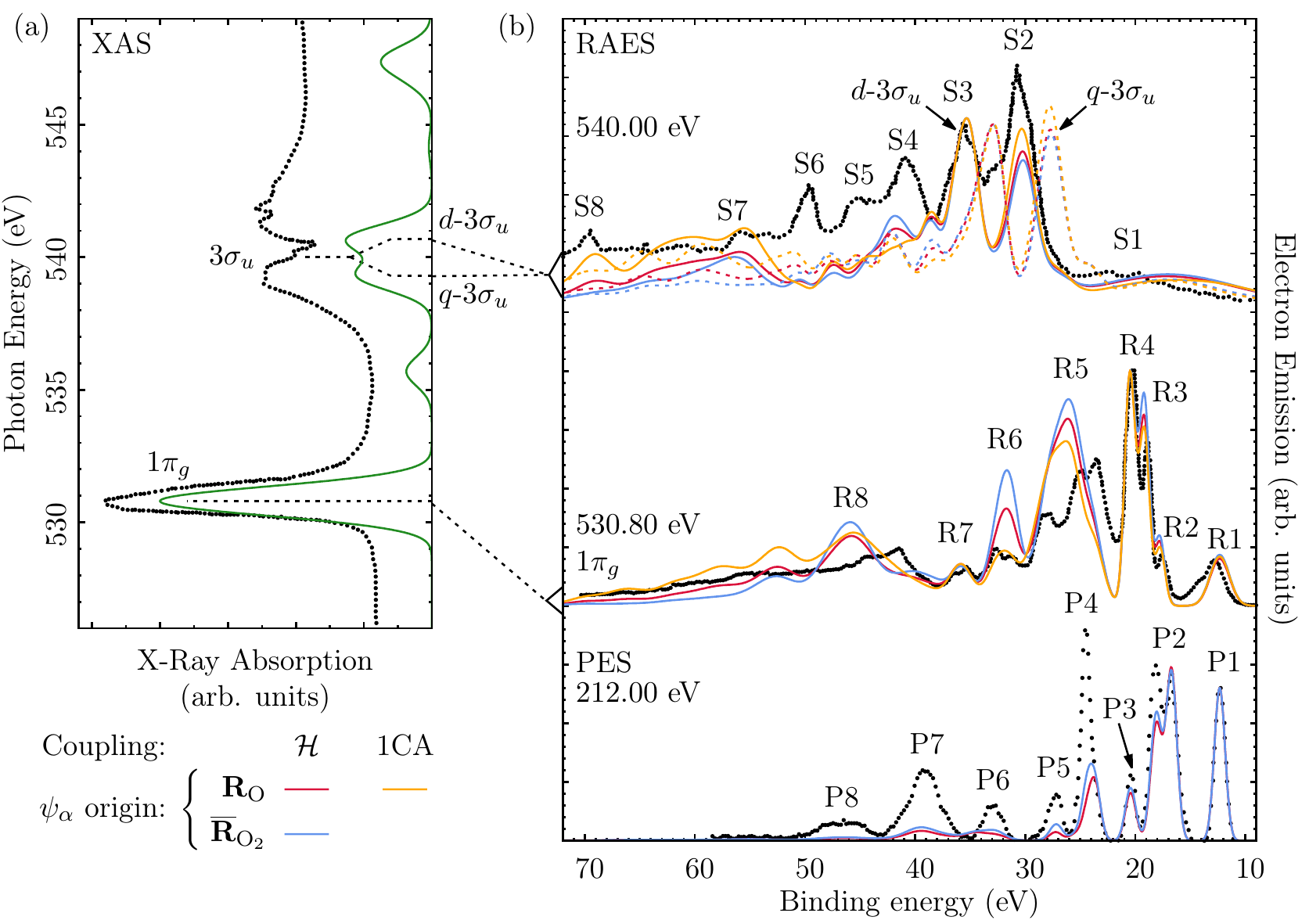}
  \caption{\label{fig:res:o2}
  (a) Calculated K-edge XAS of $\text{O}_2$ (solid) compared against experimental results from Ma et al.~\cite{Ma_PRA_1991} (dotted).
  The experimental excitation energies are connected to the theoretically found $1\pi_g$, $q\text{-}3\sigma_u$, and $d\text{-}3\sigma_u$ resonances in the \ac{XAS}, as well as to the respective \ac{RAES} in (b).
  Therein, the theoretical \ac{RAES} and \ac{PES} obtained with the \ac{SCI} approach are presented together with the respective experimental data that has been digitized from Caldwell et al~\cite{Caldwell_JESRP_1994} (dotted).
  Note that due to the high bandwidth of $\sim2.5$~eV, the whole exchange split $3\sigma_u$ double peak is excited by the incoming radiation at 540~eV.
  Hence, the theoretical \ac{RAES} for the $q\text{-}3\sigma_u$ (dashed) and $d\text{-}3\sigma_u$ (solid) resonances are depicted in this case.
  The theoretical results have been obtained with continuum orbitals corresponding to the $V_f^\text{J}(r)$ potential originating either at one atom, $\mathbf{R}_\text{O}$, or at the molecular center, $\overline{\mathbf{R}}_{\text{O}_2}$, as well as with the indicated coupling approaches.
  All spectra have been normalized.
  The applied shifts and broadenings are detailed in Table~\ref{tab:res:shift} and \supp~Table~S2, respectively.
          }
\end{figure*}
Being a biradical open shell system with triplet ground state multiplicity, molecular oxygen is an intriguing system, allowing for ionization into doublet and quartet spin states and being targeted by many studies.
Especially the core-hole decay of $\text{O}_2$ has been subject to frequent investigation, since it is a textbook example of strong lifetime-vibrational interference effects.~\cite{Sorensen_PRA_2001,Neeb_CPL_1993, Carroll_JCP_1988}
%
The plethora of available experimental data regarding its \ac{XAS}~\cite{Ma_PRA_1991,Coreno_CPL_1999,Kuiper_TJoCP_1994},
\ac{PES}~\cite{Caldwell_JESRP_1994,Sorensen_PRA_2001} and
\ac{RAES}~\cite{Caldwell_JESRP_1994, Schaphorst_CPL_1993, Kuiper_TJoCP_1994, Sorensen_PRA_2001} allows for a detailed test of our approach.
%

%
%
The homonuclear linear geometry of $\text{O}_2$ suits to test the applicability of the spherically symmetric continuum orbital model that is central to our protocol.
For instance, the $1s$ core hole might be delocalized over both atoms, whereas we expect our method to be more appropriate in the case of a localized hole.
Further, it introduces some ambiguity regarding the coordinate origin of the continuum wave functions, see Section~\ref{sec_theory}, which might be placed either on one atom, $\mathbf{R}_\text{O}$, or at the interatomic center $\overline{\mathbf{R}}_{\text{O}_2}$.
Both approaches have been tested here with respect to their ability to reproduce the \ac{PES} and \ac{RAES} of $\text{O}_2$.
%

%

\paragraph{\ac{XAS}}
\label{sec_res:o2:xas}
%
The comparison of the theoretical K-edge \ac{XAS} of $\text{O}_2$ against experimental data that has been digitized from Ma et al.~\cite{Ma_PRA_1991} is presented in Fig.~\ref{fig:res:o2} (a).
%
%
Both spectra were normalized to the dominating peak at $531$~eV that corresponds to core excitations to the $1\pi_g$ orbitals.~\cite{Ma_PRA_1991}
The structured double peak at 539-545~eV  has been assigned to two broad $3\sigma_u$ resonances that are exchange split by $2.3$~eV and a series of Rydberg excitations.~\cite{Ma_PRA_1991}
We recover this characteristic in our theory, however the splitting is smaller being about $1.4$~eV, which can be attributed to the missing Rydberg orbitals in our active space.
In agreement with Ref.~\citenum{Ma_PRA_1991}, we find the following assignment.
The lower energy feature at $539.3$~eV predominantly corresponds to the promotion of a spin-up ($\alpha$) electron to the $3\sigma_u$ orbital, i.e., quartet spin coupling in the valence shell, which we denote as $q\text{-}3\sigma_u$ in the following.
The higher energy peak at $540.7$~eV comprises doublet spin coupling in the valence shell, labeled analogously as $d\text{-}3\sigma_u$.
Further, a small feature at $535.5$~eV is not found in the experimental data, while the one at $547.5$~eV is already above the core ionization threshold.~\cite{Wurth_PRL_1990}
The former corresponds to the shake-up excitation of $1.0$ core and $0.6$ valence electrons to the $1\pi_g$ orbitals, while the latter is due to states involving simultaneous excitations to the $1\pi_g$ and $3\sigma_u$ orbitals.
The photon energy of $530.8$~eV at which Caldwell et al.~\cite{Caldwell_JESRP_1994} have recorded the \ac{RAES} has been assigned to the $1\pi_g$ resonance.
However, due to the bandwidth of $\sim 2.7$\,eV in the experimental setup, the excitation at $540.0$~eV involves both exchange-split $q\text{-}3\sigma_u$ and $d\text{-}3\sigma_u$ resonances. 
%
Notably, the evaluated core holes are $1\sigma_g^{-0.3}1\sigma_u^{-0.7}$ and $1\sigma_g^{-0.65}1\sigma_u^{-0.35}$ for the $1\pi_g$ and both $3\sigma_u$ resonances, respectively, i.e., they are delocalized in our calculation.
%
\paragraph{\ac{PES} and \ac{RAES}}
\label{sec_res:o2:raes}
%
In Fig.~\ref{fig:res:o2}(b), the computed \ac{PES} and $1\pi_g$ as well as $q\text{-}3\sigma_u$ and $d\text{-}3\sigma_u$ \ac{RAES} of $\text{O}_2$ are compared to the experimental data recorded under the pseudo magic angle of $57^\circ$ 
by Caldwell et al.~\cite{Caldwell_JESRP_1994}
The respective spectra have been normalized to the the heights of the features P1, R4, and S3.
The continuum orbitals have been generated with the $V_f^\text{J}(r)$ potential, centered either at $\mathbf{R}_\text{O}$ or
$\overline{\mathbf{R}}_{\text{O}_2}$.
Further, the partial decay rates have been estimated using the complete, $\mathcal{H}$, and approximate, $r^{-1}$, couplings.
Results obtained with the \ac{1CA} for the $\mathbf{R}_\text{O}$ origin are shown as well.
Note the remarkably different convergence of the results with respect to highest angular momentum included in the photoelectron representation: it was necessary to include waves up to $l_\text{max}=12$ for the \ac{PES}, $l_\text{max}=15$ for the \ac{RAES} with $\mathcal{H}$ and $r^{-1}$ couplings, and $l_\text{max}=5$ in case of \ac{1CA} \ac{RAES}.
%

%
%
The experimental \ac{RAES} and \ac{PES} are quite structured and feature-rich.
Remarkably, the energetic positions of experimental \ac{PES} features are reproduced almost exactly.
Using the \ac{SCI} approach we have been able to achieve quite good overall agreement with the experimental $1\pi_g$ \ac{RAES}.
Reproducing the $3\sigma_u$ spectrum is more problematic, because the $q\text{-}3\sigma_u$, $d\text{-}3\sigma_u$ and underlying Rydberg resonances,~\cite{Ma_PRA_1991} which are not included here, are simultaneously excited in the experiment.
In the \ac{PES}, P1 and P3 correspond to the ${}^2 1\pi_g^{-1}$ and ${}^23\sigma_g^{-1}$ ionic final states, respectively.
P2, having a double peak structure, is due to primarily quartet cationic states: ${}^41\pi_u^{-1}$ at $17$~eV and ${}^43\sigma_g^{-1}$ with a smaller contribution from ${}^2\pi_u^{-1}$ at $18$~eV, respectively.
P4 contains already combination transitions with the leading states ${}^21\pi_u^{-1}$ and ${}^42\sigma_u^{-0.5}3\sigma_g^{-0.5}$ which are accompanied by exciations to the $\sigma$ and $\pi$ orbitals.
P5 corresponds to the double hole states ${}^43\sigma_g^{-1}1\pi_u^{-0.75}1\pi_g^1$ and ${}^2 2\sigma_u^{-0.3} 3\sigma_g^{-0.6}1\pi_u^{-0.5}1\pi_g^{0.5}$.
%
%
%
%
%
%
%
The higher binding energy features increasingly correspond to states carrying multiple excitations that can hardly be assigned in the orbital picture.
%
%
%
Note that our analysis is in accord with that of Ref.~\citenum{Edqvist_PS_1970}.
The calculated \ac{PES} intensities quite well reproduce the experiment for P1-P3.
However, P4-P8, involving multiple excitations become increasingly underestimated with rising binding energy.
This might be due to the lack of combination transitions in this energy region because of the limited active space.
%
%
The integrated total cross sections, however, are considerably affected, being 0.30 and 0.14\,a.u. for $\mathbf R_\text{O}$ and $\overline{\mathbf R}_{\text O_2}$, respectively.
%
%

%
%
\begin{table}
  \caption{\label{tab:res:o2}
  Computed total, doublet, and quartet autoionization rates in meV for the $1\pi_g$, $q\text{-}3\sigma_u$, and $d\text{-}3\sigma_u$ resonances of $\text{O}_2$ obtained with the indicated approaches.
  The experimental reference for the total decay width of the $1\pi_g$ resonance has been taken from Ref.~\citenum{Coreno_CPL_1999}.
    }
  \begin{ruledtabular}
  \begin{tabular}{r r l l l | l l l}
    & & \multicolumn{3}{c}{$\mathbf{R}_\text{O}$} & \multicolumn{3}{c}{$\overline{\mathbf{R}}_{\text{O}_2}$} \\
    \hline
       & & $1\pi_g$ & $q\text{-}3\sigma_u$ & $d\text{-}3\sigma_u$ & $1\pi_g$ & $q\text{-}3\sigma_u$ & $d\text{-}3\sigma_u$\\
    \hline
  \multirow{3}{*}{$\mathcal{H}$}  &  Total   & $126.0$ & $125.9$ & $125.0$ & $120.2$ & $122.5$ & $121.2$ \\
                                  & Quartet  & $25.2$  & $104.8$ & $30.2$  & $35.7$  & $95.6$  & $40.6$  \\
                                  & Doublet  & $100.7$ & $21.1$  & $94.8$  & $84.5$  & $26.9$  & $80.6$  \\
    \hline
  \multirow{3}{*}{$r^{-1}$}      &  Total    & $169.0$ & $167.8$ & $165.2$ & $186.6$ & $187.9$ & $184.4$ \\
                                 & Quartet   & $48.9$  & $132.4$ & $52.8$  & $69.4$  & $140.1$ & $73.6$  \\
                                 & Doublet   & $120.1$ & $35.4$  & $112.4$ & $117.2$ & $47.8$  & $110.8$ \\
    \hline
  \multirow{3}{*}{\ac{1CA}}      &  Total    & $69.7$ & $65.1$ & $65.5$ &  &  & \\
                                 & Quartet   & $7.5$  & $57.5$ & $56.0$ &  &  & \\
                                 & Doublet   & $62.2$ & $7.6$  & $9.5$  &  &  & \\
    \hline
  Exp.                           &  Total    & $149.5\pm10$ & & & & &
  \end{tabular}
  \end{ruledtabular}
\end{table}
%


The oxygen molecule represents a convenient object to discuss the dependence of total decay rates on the continuum origin.
Moreover, it demonstrates the largest dependence of this kind among the molecules studied here and this comparison is supported by the available experimental value.
%
The total decay rates of the $1\pi_g$, $q\text{-}3\sigma_u$, and $d\text{-}3\sigma_u$ resonances obtained with all approaches are compiled in Table~\ref{tab:res:o2} together with the experimentally determined lifetime width of the $1\pi_g$ resonance.~\cite{Coreno_CPL_1999}
%
%
In the literature, it has been assumed that the $1\pi_g$ and $3\sigma_u$ resonances have approximately the same decay rate.~\cite{Schaphorst_CPL_1993, Kuiper_TJoCP_1994}
The computed total decay rates support this assumption for the results obtained within each model. 
Notably, the total decay rates of the $q\text{-}3\sigma_u$ and $d\text{-}3\sigma_u$ resonances demonstrate the expected preferential decay into quartet and doublet target states, respectively, corresponding to their valence spin coupling.
The $1\pi_g$ resonance, however, favors decay into doublet states.
Interestingly, this spin selectivity is less pronounced if the $\overline{\mathbf{R}}_{\text{O}_2}$ origin is used.
%

In absolute terms, the $\mathcal{H}$ coupling results recover approximately $80\%$ of the experimental value.~\cite{Coreno_CPL_1999}
The $r^{-1}$ coupling in turn produces total decay rates larger than the reference by $13\%$ and $25\%$ for the $\mathbf{R}_\text{O}$ and $\overline{\mathbf{R}}_{\text{O}_2}$ origins, respectively.
However, the \ac{1CA} decay rates are the smallest, providing only about $45\%$ of the experimental result.
This suggests that the \ac{1CA} misses important contributions in this case.
The delocalization of the core hole, that we discussed earlier, could be an explanation for this behavior, since the \ac{1CA} excludes all contributions from one atom.
%
Still, doubling the \ac{1CA} results recovers only about $80\%$ of the full $r^{-1}$ coupling result at the $\mathbf{R}_\text{O}$ origin, which is why we ascribe the remainder to non-local contributions.
Although the $\mathcal{H}$ and $r^{-1}$ couplings reproduce the experimental reference with similar absolute deviations, see Table~\ref{tab:res:o2}, we consider the $\mathcal{H}$ results to be in general more reliable. 
This is because our method does not include nuclear effects and therefore has less open decay channels.
Thus it is rather likely that the present model indeed should underestimate the experimentally determined total decay rate.
Further, the spectra obtained with $\mathcal{H}$ coupling agree better with the experimental $1\pi_g$ \ac{RAES} and the $r^{-1}$ results are more sensitive to the placement of the continuum origin, see \supp~Fig.~S4.
%
%

%
%
%
%
For the 1$\pi_g$ resonance (Fig.~\ref{fig:res:o2}(b)), the computed spectrum is in an overall good agreement with experiment, with the low binding energy features R1-4 being better reproduced.
This region of the spectrum  corresponds predominantly to participator decay into single hole states that constitute the \ac{PES} as well.
%
The following \ac{PES} and participator \ac{RAES} features can be assigned to the same single hole states:
P1 and R1, the high energy part of P2 and R2, P3 and R4 ($~30\%$ of participator character), and P4 constitutes the rising flank of R5.
The remainder of the \ac{RAES} is due to spectator decay to double hole and higher excited states, the representation of which might benefit from a larger active space.
For $\text{O}_2$, the \ac{1CA} \ac{RAES} agree quite well with those obtained with $\mathcal{H}$ coupling, in contrast to the behavior that we found for $\text{CH}_4$ in Fig.~\ref{fig:res:ch4}.
Further, the choice of the continuum orbital origin only weakly affects the $q\text{-}3\sigma_u$ and  $d\text{-}3\sigma_u$ spectra and the overall shape of the $1\pi_g$ one.
%
%
%
Similar to the \ac{PES} spectrum, the agreement gets worse for higher binding energies.
For instance, the intensity of R5 is overestimated, it is less structured, and its onset is shifted to higher energies by $1.5$~eV.
%
%
The atom centered approach ($\mathbf{R}_\text{O}$) performs better, {and moreover} the best agreement {especially for high-energy features} is obtained within the \ac{1CA}.
The R6 peak behaves similar, albeit the relative overestimation by the $\mathcal{H}$ coupling approaches, especially with $\overline{\mathbf{R}}_{\text{O}_2}$ is stronger.
R7, however, is reproduced well with all approaches.
Finally R8, i.e. the beginning of the tail region appears blue shifted by $2.5$~eV.
%
The intensity overestimation with the $\overline{\mathbf{R}}_{\text{O}_2}$ origin is due to the enhancement of the quartet decay branch, as indiciated by the total decay rates in Table~\ref{tab:res:o2}.
%
%

%

%
For the $3\sigma_u$ resonance the agreement with the experiment is worse than for the $1\pi_g$ resonance.
The $q\text{-}3\sigma_u$ and $d\text{-}3\sigma_u$ \ac{RAES} appear to be shifted with respect to each other both due to exchange splitting in the core-excited states and preferential decay to different spin manifolds of O$^+_2$.
%
In particular, the $q\text{-}3\sigma_u$ spectrum is shifted  to lower binding energies by $\sim2.5$~eV relative to the experiment even after the alignment of all spectra to the reference \acp{IP}, see Table~\ref{tab:res:shift}.
Further, the intensity of the features with a binding energy above 35\,eV (S4-6) are considerably underestimated. 
This discrepancy may be attributed mainly to two reasons as discussed previously in experimental works.~\cite{Schaphorst_CPL_1993, Kuiper_TJoCP_1994,Ma_PRA_1991,Lapiano-Smith_TJoCP_1990}
As mentioned before, the $3\sigma_u$ band is overlaid by a manifold of Rydberg transitions in addition to the exchange splitting between $q\text{-}3\sigma_u$ and $d\text{-}3\sigma_u$ resonances.~\cite{Ma_PRA_1991}
Namely, within the excitation bandwidth of $2.7$~eV,~\cite{Caldwell_JESRP_1994} Rydberg states of $np(\pi_u) , n=3\text{--}7$ and $ns(\sigma_g), n=3\text{--}5$ character 
have been identified by Ma et al.~\cite{Ma_PRA_1991} around $540$~eV.
This indicates that the measured spectrum contains contributions from all these resonances and in our calculation we consider only one at a time.
Further, the $3\sigma_u$ states are dissociative and a fast bond elongation is expected to occur on the timescale of the Auger decay 
that has been claimed to lead to appearence of atomic features.~\cite{Kuiper_TJoCP_1994, Schaphorst_CPL_1993, Caldwell_JESRP_1994}
%
Our test calculations with a bond distance of $1.26$~{\AA} ($0.05$~{\AA} elongation) show for $q\text{-}3\sigma_u$ an almost doubled energetic mismatch between theory and experiment of $4.5$~eV (compare with 2.5\,eV), indicating that a simple bond elongation might not explain the effect.
%
%
To close this question, the one-step model for resonant Auger decay~\cite{Aberg_CaRiMI_1982} has to be employed together with a larger active space that includes excitations to Rydberg orbitals.
Finally, varying the excitation ratio of all these resonances might strongly influence the result.

%
Summarizing, the total Auger decay rates, the $1\pi_g$ \ac{RAES} as well as the \ac{PES} are in quite good agreement with the experiment, whereas the $3\sigma_u$ \ac{RAES} agrees slightly worse calling for more sophisticated treatment.
%
%
In general, the full $\mathcal{H}$ coupling seems to yield the most stable spectra and total decay rates and the \ac{1CA} predicts spectra of the same quality as the full approaches, although the overall decay rates are underestimated.
Finally, the choice of the continuum origin has no strong effect on the spectra and total decay rates, although the total photoionization cross sections vary by a factor of two.
%
%

\subsection{Nitrogen dioxide}
\label{sec_res:no2}
%
\begin{figure*}
  \includegraphics{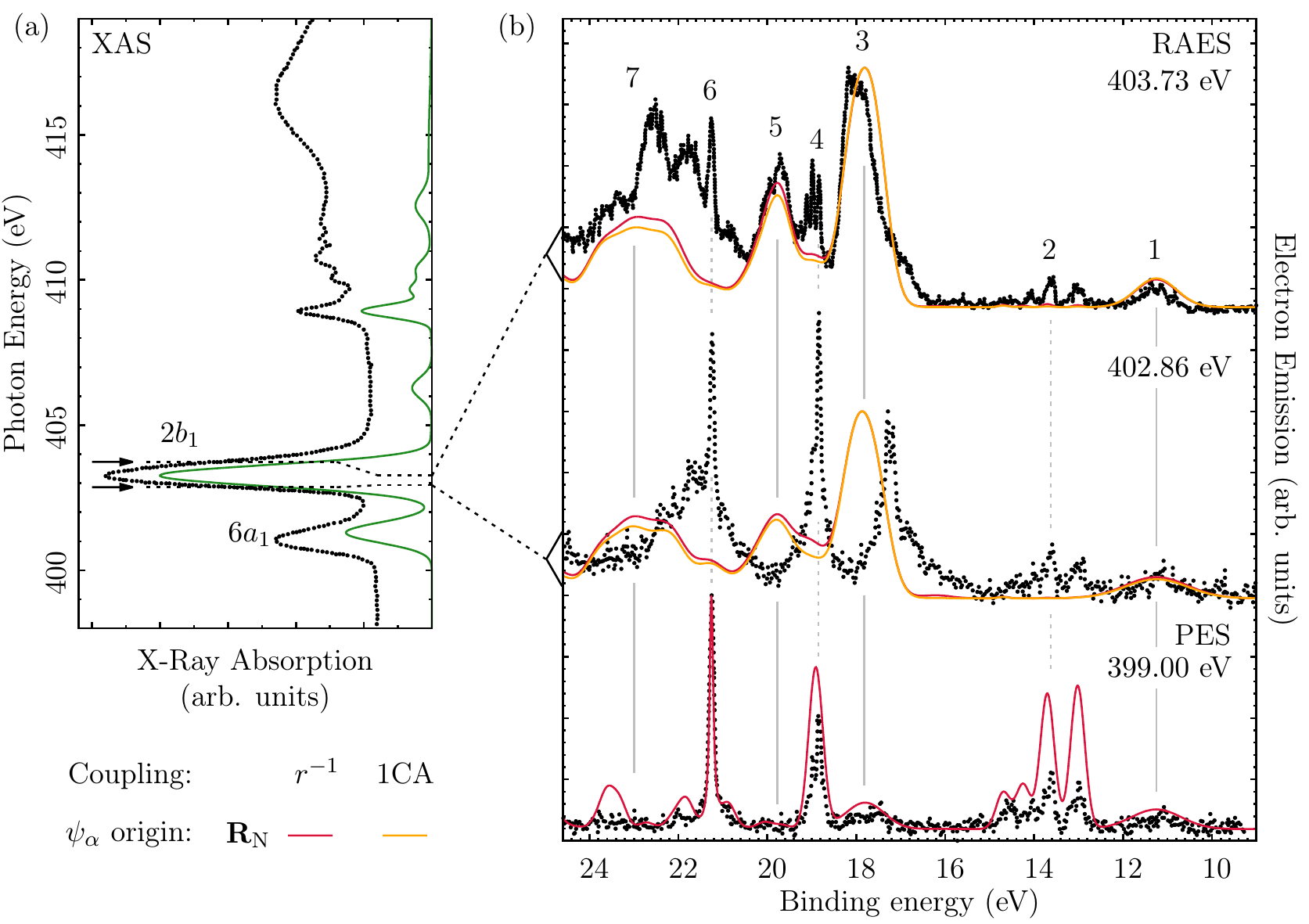}
  \caption{\label{fig:res:no2}
  (a) Theoretical nitrogen K-edge \ac{XAS} of $\text{NO}_2$ (green, solid) and the  experimental reference digitized from Gejo et al.~\cite{Gejo_CP_2003} (black, dotted).
  %
  %
  The arrows indicate the excitation energies used by Piancastelli et al.~\cite{Piancastelli_CPL_2004} to obtain the experimental \ac{RAES} in panel (b).
  In addition, the dashed lines link the experimental excitations to the respective core-excited states predicted in our calculation and the corresponding \ac{RAES} in (b).
  Therein, the decay spectra measured at the low and high energy sides of the $2b_1$ resonance as well as the \ac{PES} taken at $399$~eV below the edge~\cite{Piancastelli_CPL_2004} (dotted) are compared against our theoretical results (solid).
  %
  %
  The partial decay rates have been obtained using $r^{-1}$ coupling and the \ac{1CA}.
  The vertical lines indicate corresponding features that are predominantly found in \ac{PES} (dashed) and \ac{RAES} (solid), respectively.
  %
  %
  The applied shifts and broadenings are detailed in Table~\ref{tab:res:shift} and in \supp~Table~S3.
  }
\end{figure*}
$\text{NO}_2$, being a radical with multi-configurational character of the ground doublet state,
represents an increase of complexity with respect to $\text{O}_2$.
This fact makes it a convenient object to study the performance of our multi-reference protocol.
Moreover, recent reference data for its \ac{XAS}~\cite{Jurgensen_CP_2000, Gejo_CP_2003},
\ac{PES}~\cite{Piancastelli_CPL_2004}, and \ac{RAES}~\cite{Piancastelli_CPL_2004} taken at the nitrogen and oxygen K-edges are available for comparison.
In this article, we focus on the nitrogen K-edge.
This takes advantage of the fact that the origin of the spherically symmetric continuum functions can be unambiguously put on the nitrogen atom $\mathbf{R}_\text{N}$.
%
%
%

%
\paragraph{XAS}
\label{sec_res:no2:xas}
%
The calculated \ac{XAS} at the nitrogen K-edge of $\text{NO}_2$ 
is presented in Fig.~\ref{fig:res:no2} panel (a), together with the experimental reference that has been digitized from Gejo et al.~\cite{Gejo_CP_2003}
%
%
Both spectra have been normalized to the height of the $2b_1$ peak.
%
%
%
The relative intensities of the $6a_1$ and $2b_1$ resonances are reproduced almost exactly, although the $6a_1$ peak is predicted slightly by $0.2$~eV too high in energy.
The features at higher energies correspond to excitations to the $7a_1$, $5b_2$, and Rydberg orbitals~\cite{Gejo_CP_2003} that are not included in our active space.
Thus, they are reproduced herein only to some extent.
The excitation energies of $402.86$~eV and $403.73$~eV used to obtain the \ac{RAES} from the low and high energy flank of the $2b_1$ resonance~\cite{Piancastelli_CPL_2004} are indicated by the arrows.
We have connected them to the positions of two core-excited states predicted by our calculations that correspond to the $2b_1$ resonance.
Our calculation predicts a splitting of $0.34$~eV for these states.
Further, the oscillator strength obtained for the lower state is a factor of $\sim20$ smaller than the one of the higher state.
An analysis in terms of the occupation numbers of the state averaged orbitals obtained from the \ac{CASSCF} procedure shows that both states are quite similar in character, bearing the $\text{N}(1s)$ core hole and the excitation to the $2b_1$ orbital.
%
However, the analysis of the core-hole spin in our calculations does not support the assignment of this splitting to an exchange mechanism, in contrast to the experimental results for the $\text{O}(1s)\rightarrow2b_1$ resonance.~\cite{Piancastelli_CPL_2004}
%
%
%
%
%
\paragraph{PES and RAES}
\label{sec_res:no2:raes}

Panel (b) comprises the \ac{RAES} corresponding to the low and high energy flanks of the $2b_1$ resonance, as well as the direct \ac{PES} taken at $399$~eV below the edge.
The experimental data has been digitized from Ref.~\citenum{Piancastelli_CPL_2004}.
The \ac{PES} and \ac{RAES} have been normalized to the peaks 4 and 3, respectively.
%
To converge the partial decay rates and the ionization cross sections it was necessary to include partial waves up to $l_\text{max}=9$ ($r^{-1}$ coupling) and $l_\text{max}=4$ (\ac{1CA}) as well as $l_\text{max}=15$ (\ac{PES}), respectively.
%
%
Note that we have used a Lorentzian lineshape with an \ac{FWHM} of $0.1$~eV for peak 6 in the calculated \ac{PES}.
To have a common reference, the theoretical and experimental data have been aligned to reference values for the lowest singlet and triplet \acp{IP} as detailed in Table~\ref{tab:res:shift}.
For the sake of clarity a common set of identifiers 1-7 is used for features in all spectra.
The dashed lines denote features or groups thereof that predominantly correspond to direct photoionization, whereas those that are found in both, \ac{PES} and \ac{RAES}, are indicated with solid lines.
%

\ac{PES} is found to be in fairly good agreement with experiment,
although the feature 2 is overestimated by an approximate factor of 2.
	Further, the peak 4 is slightly overestimated and around $23.6$~eV our simulation yields a feature that has no correspondence in the experimental data.
	Some of this deviations might be due to anisotropy effects present in the experimental data~\cite{Piancastelli_CPL_2004}  
	which are not recovered in our angle integrated spectra.
	In addition, the valence \ac{PES} is a rather non-local probe and thus might be more sensitive to the quality of the continuum orbital than \ac{RAES}, which in this case seems to be an almost purely local process.
	An improvement could thus require a more elaborate model for the continuum electron, possibly involving a multi-centered approach.~\cite{Marante_JCTC_2017}

We address here different flanks of the $2b_1$ resonance which exhibits an exchange splitting of $\sim0.5$~eV for the $\text{O}(1s)$ hole~\cite{Piancastelli_CPL_2004, Gejo_CP_2003}.
Piancastelli et al.~\cite{Piancastelli_CPL_2004} found that the low energy side of the $\text{O}(1s)\rightarrow 2b_1$ resonance favors decay to triplet ionized states whereas the higher energy side favors decay to singlet states, which is a trace of this splitting.
A similar behavior can be expected for the  $\text{N}(1s)\rightarrow 2b_1$ resonance, although the splitting is assumed to be smaller~\cite{Piancastelli_CPL_2004}.
However, to the best of our knowledge no data regarding the splitting has been published yet and the experimental \ac{RAES} results in Ref.~\citenum{Piancastelli_CPL_2004} are inconclusive in this case.
We attempt to shed light on this question in the following.

The results of the $r^{-1}$ coupling are presented in Fig.~\ref{fig:res:no2} assuming the full molecular treatment and \ac{1CA}.
Noteworthy, the \ac{RAES} obtained within the \ac{1CA} closely resemble the ones that have been obtained by taking into account non-local contributions from the oxygen atoms as well.
%
Adding to that, the calculated total decay widths of $92$~meV and $98$~meV corresponding to the \ac{1CA} and the full $r^{-1}$ couplings agree with each other as opposed to oxygen case.
%
Thus it seems that the decay of the $1b_2$ resonance can be described quite well as a purely local phenomenon, in contrast to our findings for $\text{CH}_4$ and $\text{O}_2$.
To unravel the relative contribution of the triplet and singlet decay pathways, 
we found T:S ratios of $1.0$ and $1.2$ for the the low and high energy $1b_2$ \ac{RAES}.
%
Thus there is no appreciable difference in the spin coupling of the respective core-excited states, supporting the result from the analysis of the respective core-hole spins, see the discussion of the \ac{XAS} above.
In contrast, the total photoionization cross section yields a T:S ratio of $2.5$, i.e., the \ac{PES} at this energy is dominated by the triplet branch.
%
This behavior fits well with the fact that the $6a_1$ orbital is occupied in the $\text{NO}_2$ ground state, as well as in a large part of the triplet ionized states but only to lesser extent in the singlet states.
Concerning the analysis of the spectra, the comparison against the reference data~\cite{Piancastelli_CPL_2004} yields a very good agreement regarding the energetic positions of all features in the high energy \ac{RAES} and the \ac{PES}.
%
%
The peaks 2, 4, and 6 are predominant \ac{PES} features and their presence in the experimental \ac{RAES} indicates that a considerable portion of the absorbed photon flux leads to a direct ionization rather than to a resonant excitation of the molecule.
To reproduce these effects theoretically, the one-step model~\cite{Aberg_CaRiMI_1982} for resonant Auger decay must be employed.
%

%
The frontier orbital configurations for $\text{NO}_2$ as well as singlet and triplet $\text{NO}_2^+$ are $4b_2^2 1a_2^2  6a_1^1$, $4b_2^2 a_2^2 6a_1^0$, and $1a_2^2 4b_2^1 6a_1^1$, respectively.
The $2b_1$ orbital however, is unoccupied in all ground states with given charge and multiplicity.
Hence, core excitations to the $2b_1$ orbital can be expected to have a considerable participator decay contribution.
In contrast, the peaks 3, 5, and 7 are predominant spectator Auger decay contributions having only small counterparts in the \ac{PES}.
Finally, the lowest energy peak, 1, corresponds to the singlet $\text{NO}_2^+$ ground state, i.e. the ${}^1 6a_1^{-1}$ ionization.
It can be reached via direct ionization and participator decay of the $2b_1$ resonance and contributes equally to both, \ac{PES} and \ac{RAES}.
In opposite to that, participator decay into the higher lying single hole states contained in feature 2, i.e., ${}^3 4b_2^{-1}$, ${}^3 1a_2^{-1}$, ${}^1 1a_2^{-1}$, and ${}^1 4b_2^{-1}$ at $13$~eV, $13.6$~eV, $14.2$~eV, and $14.6$~eV, is suppressed, although this states contribute to the \ac{PES}.
We can confirm the statement of Ref.~\citenum{Piancastelli_CPL_2004} that this is most likely due to the localization of the $1a_2$ and $4b_2$ orbitals on the oxygen atoms, whereas the $2b_1$ excitation is primarily localized on the nitrogen. 
Hence, the main part of the Auger decay of the $2b_1$ resonance at the N K-edge of $\text{NO}_2$ can be ascribed to spectator decay into double-hole or higher excited states, which appear only as satellites in the \ac{PES}.
This is a common situation and underlines the importance of the joint analysis of \ac{PES} and \ac{RAES}. 
%
The strong configuration mixing in the electronic structure calculation hinders a clear assignment of the remaining features, thus, we refer to the assignments in Refs.~\citenum{Schirmer_CP_1981} and~\citenum{Katsumata_CP_1982}.
%

One can expect the Auger features to shift on the binding energy scale when changing the photon energy, the energetic position of the direct ionization \ac{PES} features, however, should stay intact.
The theoretically obtained \ac{RAES} taken at the low-energy flank is rather similar to our high-energy result and does not recover the shift of the decay features 3, 5, and 7 by approximately $0.6$~eV observed in the experimental spectrum.
Judging by the relative contribution of the direct ionization (\ac{PES}) features 2, 4, and 6 to the measured data, the resonant decay contribution is here much weaker than in the \ac{RAES} taken at high-energy flank.
%
In turn, feature 1 can be ascribed exclusively to direct photoionization that does not appear red-shifted in contrast to its participator decay counterpart that could be concealed within the noise due to its small intensity.
Since our preliminary studies with a larger active space and basis set have not led to an improvement here, we tend to attribute this shift to difference in exciting photon energies and vibrational effects.
%
Nevertheless, we do not exclude other possible explanations. 
%

%

%
Finally, concerning the relative intensities.
The overall agreement is good, with some exclusions.
For instance in the high energy flank \ac{RAES}, the relative intensities of the features 1, 3, and 5 are reproduced quite well, while the peak 7 is underestimated considerably and is less structured than in experiment.
%
%
The situation is more delicate concerning the low energy \ac{RAES}.
In this case, the overall small magnitude of the decay features in the experimental spectrum could be due to the low oscillator strength of the corresponding core-excited state, as discussed in the \ac{XAS} part.
Further, the red shift of the resonant decay features by $0.6$~eV increases the overlap between the direct photoionization peaks 4 and 6 with the decay features 5 and 7, respectively.
Thus a disentanglement of direct and resonant contributions is hardly possible for these peaks.
It seems that for a better reproduction of the experimental results for the low energy \ac{RAES}, the one-step ansatz to \ac{RAES}, allowing for interference between  multiple resonant decay and direct ionization channels~\cite{Aberg_CaRiMI_1982} as well as the inclusion of vibrational effects have to be considered.
%

%

%
%
In summary, the comparison to the experimental \ac{XAS} has verified that our electronic structure is suitable to describe the $2b_1$ resonance at the nitrogen K-edge.
Further, our theoretical data agree well with the experimental reference for the \ac{PES} and the \ac{RAES} recorded at the high energy flank of the resonance.
In addition, the \ac{1CA} results closely resemble the spectra and total decay rates obtained when all atomic centers are included.
However, the \ac{RAES} stemming from the low energy side of the resonance probably requires a more involved treatment of the resonance decay within the one-step model as well as the inclusion of nuclear effects.
Finally, our results do not indicate a difference in the spin coupling on the low and high energy flank of the $\text{N}(1s)\rightarrow2b_1$ resonance, in contrast to previous findings for the $\text{O}(1s)\rightarrow2b_1$ resonance.~\cite{Piancastelli_CPL_2004}
%
\subsection{Pyrimidine}
\label{sec_res:pyr}
%
\begin{figure*}
  \includegraphics{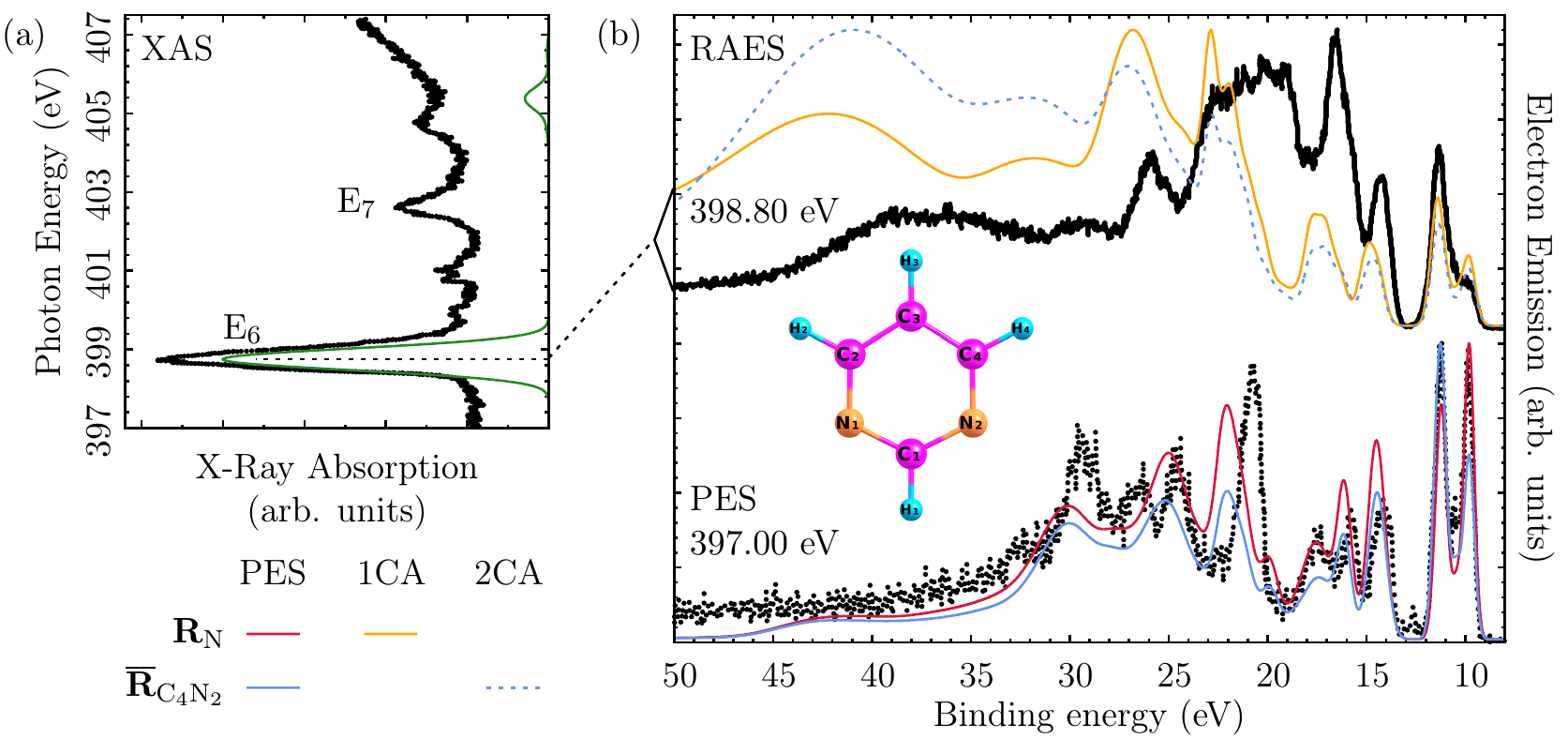}
  \caption{\label{fig:res:pyr}
  (a) Theoretical nitrogen K-edge \ac{XAS} of pyrimidie (solid green) and the respective experimental measurement taken from Bolognesi et al.~\cite{Bolognesi_JCP_2012} (black dotted).
  (b) Theoretical and experimental~\cite{Bolognesi_JCP_2012} (black) \ac{RAES} of the E6 resonance and the pre-edge \ac{PES} obtained at the indicated energies.
  The continuum orbitals have been evealuated using the $V_f^\text{J}(r)$ potential centered at one nitrogen atom, $\mathbf{R}_\text{N}$, or at the geometric center of the ring, $\overline{\mathbf{R}}_{\text{C}_4\text{N}_2}$.
  The \ac{RAES} have been evaluated in the \ac{1CA} and \acs{2CA}, see text.
  The applied shifts and broadenings are detailed in Table~\ref{tab:res:shift} and \supp~Table~S4, respectively.
  }
\end{figure*}
Finally, we benchmark the \ac{SCI} approach for the pyrimidine molecule ($\text{C}_4\text{H}_4\text{N}_2$).
Having six heavy atoms arranged in a $C_{2v}$ symmetric ring, the theoretical study of its \ac{PES} and \ac{RAES} is considerably more involved than for the previous examples.
Moreover, pyrimidine is interesting on its own, since the structures of the nucleic acids uracil, cytosine, and thymine are derived from it.
Its ionization spectra are thus an important landmark on the route towards a better understanding of DNA radiation damage.
Due to this, a number of experimental and theoretical investigations regarding its \ac{XAS}~\cite{Vall-llosera_TJoCP_2008, Bolognesi_JCP_2012, Bolognesi_JCP_2012},
\ac{PES}~\cite{Holland_CP_2011, Bolognesi_JCP_2012, OKeeffe_MP_2009,Potts_JPBAMOP_2003, Lottermoser_EJOC_2005},
 normal~\cite{Storchi_TJoCP_2008} and resonant~\cite{Bolognesi_JCP_2012} Auger emission have been published during the last 20 years.
Bolognesi et al.~\cite{Bolognesi_JCP_2012} have recently reported complete sets of \ac{XAS}, \ac{PES}, and \ac{RAES} measurements at the carbon and nitrogen K-edges of pyrimidine, making this work a suitable reference for our simulations.
Conveniently, the electrons have been collected at the pseudo magic angle, suppressing angular anisotropy effects.
We have chosen to focus on the nitrogen K-edge of pyrimidine, since the corresponding \ac{RAES} are quite structured and it allows to incorporate only two instead of four core orbitals into the active space.
However, the symmetric arrangement of the nitrogen atoms, Fig.~\ref{fig:res:pyr}, might lead to core-hole delocalization effects.
Similar to the $\text{O}_2$ case, see Sec.~\ref{sec_res:o2}, the continuum orbital origin is also not well defined.
Here, two options seem reasonable.
First, it might be placed on one of the nitrogen atoms, denoted as $\mathbf{R}_\text{N}$, or 
%
in the geometric center of the ring, $\overline{\mathbf{R}}_{\text{C}_4\text{N}_2}$, which might be the most balanced choice if all atoms contribute to the continuum orbital. 
%

%
%
%
%
%
%
%
%
%

\paragraph{XAS}
\label{sec_res:pyr:xas}
%
The calculated nitrogen K-edge \ac{XAS} of pyrimidine 
is depicted in panel (a) of Fig.~\ref{fig:res:pyr} together with the experimental reference;~\cite{Bolognesi_JCP_2012}
%
%
the band labels are taken from this experiment.
In case of pyrimidine, the agreement with the experimental \ac{XAS} is less satisfactory than for the other molecules.
%
%
In accord with previous findings,~\cite{Bolognesi_JCP_2012, Bolognesi_TJoCP_2010} we assign the main features to a $\pi^*(a_2)$ core excitation (E6) and to transitions to the $\pi$ - Rydberg mixed $\pi^\text{Ryd}(b_1)$ orbitals ($\sim406$~eV; E7).
In contrast, the shoulder of E6 at $399.9$~eV that has been attributed to $\pi^*(b_1)$ excitations~\cite{Bolognesi_TJoCP_2010} is missing in our results.
In fact, the corresponding core-excited state is present at $400.4$~eV, but its absorption is by a factor of $\sim500$ lower than for the E6 one.
A different theoretical study,~\cite{Vall-llosera_TJoCP_2008} however, found this feature to be underestimated as well, but perceivable.
Finally, the structure around $401$~eV has been attributed to a contamination with $\text{N}_2$ molecules~\cite{Bolognesi_TJoCP_2010} and can be neglected in the discussion.
The energetic spacing between E6 and E7 is overestimated by about 2.8\,eV in our calculations. 
This is an indication of the lack of electron correlation as well as Rydberg character in the active orbitals.
However, including a higher level of \ac{CI} and employing much larger Rydberg bases has not been feasible.
%

%
\paragraph{\ac{PES} and \ac{RAES}}
\label{sec_res:pyr:raes}
%

The predicted \ac{PES} at $397$~eV and the \ac{RAES} obtained at the E6 resonance together with experimental data digitized from Bolognesi et al.~\cite{Bolognesi_JCP_2012} are compiled in Fig.~\ref{fig:res:pyr} (b).
All spectra have been normalized to their respective highest peaks.
We focus here solely on the E6 resonance, because its spectrum yields more structure than the E7 one~\cite{Bolognesi_JCP_2012} and we deem our core-excited states to be more reliable for this lower lying resonance.
The continuum orbitals used in the calculations have been constructed based on the $V_f^\text{J}(r)$ potential centered at the $\mathbf{R}_\text{N}$ and $\overline{\mathbf{R}}_{\text{C}_4\text{N}_2}$ origins, as depicted in Fig.~\ref{fig:res:pyr}.
%
%
%
Due to the computational costs, the \ac{RAES} have only been obtained in the nitrogen centered \ac{1CA}, as well as the \ac{2CA}.
For the latter spectrum contributions from at most two different atoms of the $\mathrm{C_4N_2}$ ring have been included in a two-center approximate fashion with the continuum orbitals being centered at $\overline{\mathbf{R}}_{\text{C}_4\text{N}_2}$.
We have included partial waves up to $l_\text{max}=15$ for the \ac{PES} and \ac{2CA} \ac{RAES}, which is the current maximum supported by our code.
However, the contributions of the $l=15$ term can for some transitions reach up to $15\%$ and $25\%$ for the $\overline{\mathbf{R}}_{\text{C}_4\text{N}_2}$ (\ac{PES} and \ac{2CA} \ac{RAES}) and $\mathbf{R}_\text{N}$ (\ac{PES}) origins, respectively.
Hence, partial waves with higher angular momenta might be needed to ensure complete convergence of the respective band intensities.
%
However, for the \ac{1CA} decay spectrum, corresponding to the autoionization contribution from one nitrogen atom embedded in the molecular potential, the intensities converged already for $l=4$, although $l_\text{max}=9$ has been used.
%

%
%
%

%
At first glance, our simulations reproduce the experimental \ac{PES} results very well, whereas the \ac{RAES} intensities are matched to a lesser extent.
%
%
For the detailed discussion, we follow Bolognesi et al.~\cite{Bolognesi_JCP_2012} in dividing the spectra into three regions, spanning binding energies up to $13$~eV (1), $20$~eV (2), and $>20$~eV (3).
We begin the discussion with the \ac{PES}.
%
%
Region (1) contains clearly distinguishable single hole states with the leading configurations $n_{\text{N}_-}(7b_2)^{-1}$ at $9.8$~eV, $\pi_3(2b_1)^{-1}$ at $10.5$~eV, $n_{\text{N}_+}(11a_1)^{-1}$ at $11.2$~eV and $\pi_2(1a_2)^{-1}$ at $11.5$~eV
in Fig.~\ref{fig:res:pyr} (b).
This assignment agrees with the one obtained in Refs.~\citenum{OKeeffe_MP_2009,Holland_CP_2011}, regarding the still controversial question of the ordering of the $n_{\text{N}_+}(11a_1)^{-1}$ and $\pi_2(1a_2)^{-1}$ features~\cite{Bolognesi_JCP_2012} that are difficult to resolve experimentally.
The region (2), marks the transition from single to double hole states.
The feature around  $14.5$~eV comprises three states of dominant $\pi_1(1b_1)^{-1}$, $\sigma(10a_1)^{-1}$ and $\sigma(6b_2)$ nature with an admixture of up to $20\%$ $\pi^*$ character.
The other two peaks, correspond dominantly to lower lying $\sigma^{-1}$ states with up to $30\%$ $\pi^*$ character.
Finally, in region (3), the density of the ionized states increases considerably and they are predominantly of two hole character.
Here, multiple transitions contribute to each peak and an assignment in terms of orbitals is not possible anymore.
%

%
%
Concerning the intensities, in region (1), the height ratio of the double peak is not symmetric.
The low and high energy parts are enhanced with the $\mathbf{R}_\text{N}$ and $\overline{\mathbf{R}}_{\text{C}_4\text{N}_2}$ origins, respectively.
%
%
In region (2), $\overline{\mathbf{R}}_{\text{C}_4\text{N}_2}$ leads to a better agreement for the first two peaks, whereas using $\mathbf{R}_\text{N}$ resembles the part around $17.5$~eV better.
The experimental data in region (3) yields four distinct features around $21$~eV, $25$~eV, $26.5$~eV, and $29.5$~eV as well as an unresolved tail at higher energies.
Our data reproduces the general structure, albeit the features appear blue shifted by $1.3$~eV, $0.5$~eV, $1.1$~eV, and $0.8$~eV, respectively.
Further, the relative intensity of this region tends to be underestimated, in particular with the $\overline{\mathbf{R}}_{\text{C}_4\text{N}_2}$ origin.
Especially the feature around $27.5$~eV appears only as a small shoulder.
Since the \ac{PES} of $\text{O}_2$ demonstrated only minor variations for different origins, one might attribute the deviations observed here to the fact that the angular momentum expansion is not fully converged in this case.
%
%

%

%
Now concerning the \ac{RAES} of the E6 resonance, Fig.~\ref{fig:res:pyr} (b).
%
%
Both, the \ac{1CA} and \ac{2CA} spectra show only a very approximate correspondence to the experimental data.
%
%
In region (1) the theoretical \ac{RAES} resembles the shape of the experimental reference.
It can be assigned to pure participator decay into the single hole states that constitute the \ac{PES} in this region as well.
%
Already in region (2), we find that spectator decay is dominating 
and the number of final states increases to around 40 for the whole region. 
%
%
Here, our data yields two distinct peaks similar to the experimental result, but blue shifted by about $0.7$~eV.
The relative intensity of the one at $17.5$~eV, which is the most prominent feature in the experimental data, however, is underestimated by a factor of $2.5$ in the \ac{1CA}.
Finally, region (3) comprises more than $250$ mainly spectator transitions with appreciable intensities that in total constitute the spectrum.
Here, the experimental data comprises 4 features.
A broad one ($19.0-24.0$~eV) is only partially reproduced in the theoretical spectrum.
It is too narrow, spanning only $21.5-23.5$~eV and its rising edge is blue shifted by approximately $3.0$~eV.
Next, a narrower peak around $26.0$~eV is found in our data with a blue shift of $0.5$~eV but too large in intensity.
Furthermore, the two broad features around $29.0$~eV and $37.0$~eV in the experiment are represented with respective blue shifts of $3.0$~eV and $5.0$~eV.
Generally, the tail region, $>25.0$~eV carries too much intensity in our calculation, especially in the \ac{2CA} variant.
In comparison to our \ac{RAES} results for $\text{O}_2$ and $\text{NO}_2$, we find the extent of the disagreement between the theoretical and experimental decay spectra surprising.
This is especially striking, since the \ac{PES} is represented quite well by our approach, indicating that the valence electronic structure of the cation is well described with the present active space and that the employed continuum model captures the most important effects.
Unfortunately, a detailed scrutinization of this deviation is difficult since, to the best of our knowledge, references obtained on a higher level of theory are not available.
%
%
Tentatively, the disagreement can be assigned to the exclusion of the hydrogen atoms within both, \ac{1CA} and \ac{2CA}, which might induce a similar characteristic as for methane, see Fig.~\ref{fig:res:ch4}.
This line of argumentation is supported by the fact that region (2), with the greatest mismatch to the experimental reference, comprises ionized states bearing holes in the $\sigma$ valence orbitals which contain a considerable hydrogen character.
Further, the \ac{2CA} ansatz, that includes all ``heavy'' atoms does not improve the agreement with the experimental data in this region, although the total decay rates increase by a factor of 3.5 from 32.5~meV (\ac{1CA}) to 114.7~meV (\ac{2CA}).
While this indicates that non-local contributions from the C and N atoms play a considerable role for the decay of the E6 resonance, it also shows that the inclusion of the hydrogen atoms might be necessary, to further improve the spectrum.
%
%
Another reason could be that the active space used herein does in fact not describe the core-excited states well enough, as pointed out in the discussion of the \ac{XAS}.
We conclude the discussion of this system with the following recapitulation.
The pyrimidine molecule has been studied with a large active space, allowing one electron to be removed from the $\text{N(1s)}$ orbitals, as well as a single electron to be excited into $8$ valence orbitals.
With this active space not all prominent \ac{XAS} features can be be reproduced.
However, irrespective of the continuum orbital origin, the experimental \ac{PES} has been reproduced quite well with the \ac{SCI} approach, although it is not fully converged at $l_\text{max}=15$.
This also indicates that the valence electronic structure is described sufficiently accurate.
Still, the simulated \ac{RAES} of the E6 resonance obtained with our approach yield notable differences to the experimental reference~\cite{Bolognesi_JCP_2012} in the spectator decay region.
We tentatively ascribe this to the exclusion of the hydrogen atoms within the employed \ac{1CA} and \ac{2CA} schemes, which might influence different spectral regions to a varying degree, thus altering the shape of the spectrum.
Further studies including the hydrogen orbitals as well as involving an active space that covers more core-hole correlation need to be carried out to rule out these remaining uncertainties.
%
\section{Conclusions and Outlook}
\label{sec_conclusion}
%
In this article, we have presented the \ac{SCI} approach for the evaluation of photoionization cross sections and partial Auger decay rates for the case of molecules.
It is a logical continuation of our previous benchmark of the protocol for the atomic case of \ac{RAES} of the neon $1s^{-1}3p$ resonance, where it was shown to yield spectra and total decay rates in good agreement with experimental references.~\cite{Grell_PRA_2019}
%
The central approximation in the protocol is that the angular structure of the molecular potential is averaged out, leading to spherically symmetric continuum orbitals that are obtained by numerically solving the radial Schr\"odinger equation.
%
%
As has been shown here, such an approach, being natural for atoms, in fact provides a valuable insight into the nature of  molecular photoionization and autoionization spectral features as well.

The investigated molecules have been selected to represent different classes.
To perform a thorough test of the protocol, the \ac{XAS}, \ac{PES}, and \ac{RAES} at the carbon, oxygen and nitrogen K-edges of $\mathrm{CH_4}$, $\text{O}_2$, and $\mathrm{NO_2}$ as well as pyrimidine, respectively, have been evaluated and compared to the experimental data.
Of course, accurate calculations of these molecules in the gas phase require the inclusion of vibrational effects, with  the O$_2$ $3\sigma_u$ resonance being an extreme case.
Here, they have been excluded and the discussion is focused on the purely electronic effects.
%

%


From the viewpoint of electronic structure calculations, this protocol can be applied together with any quantum chemistry method, allowing for a \ac{CI}-like representation of the wave function.
In particular, in the present article the bound electronic structure is obtained at the \ac{RASSCF}/\ac{RASPT2} level.
The analysis of the \ac{XAS} has shown that this method with the respective active spaces is capable of capturing the most important effects in core-excited states of neutral systems with the exception of pyrimidine, possibly requiring a larger active space.
%
%
Further, the energetic positions of lines in \ac{PES} and \ac{RAES}, characterizing the structure of the valence levels of the ionized system
%
could be reproduced with an overall high accuracy.
However, there is a tendency to overestimate the binding energies towards high-energy parts of the spectra.

%
%

%
Regarding the \ac{PES} intensities, the \ac{SCI} approach has lead to fairly good agreement with experiments for $\mathrm{NO_2}$, pyrimidine,
and the lower binding energy region of $\text{O}_2$, but in some cases discrepancies have been observed.
%
%
For example, the high binding energy region of the $\text{O}_2$ \ac{PES} bears too few intensity, which we attribute to the lack of combination transitions to Rydberg orbitals in the active space.
In $\mathrm{CH_4}$, the relative intensities of the $a_1^{-1}$ and $t_2^{-1}$ peaks  \ac{PES} at 282.2~eV can not be correctly described using our approach.
%
%

The \ac{SCI} approach appears to produce in general reasonable \ac{RAES}, for all resonances in methane, the $1\pi_g$ resonance of O$_2$, and the high-energy flank of the $2b_1$ resonance in NO$_2$.
Those spectra which cannot be accurately reproduced generally correspond to cases when either vibronic effects are important, several resonances are overlaying, or the active space is too small.
Nevertheless, this information can be still valuable for the assignment of experimental spectra.

The $\mathcal H$ coupling has been applied only to CH$_4$ and O$_2$, where it has led to the best agreement.
However, the approximate $r^{-1}$ coupling produces very similar spectra for CH$_4$, whereas for O$_2$ the differences between both couplings are larger.
Overall, the origin dependence of the spectra in both O$_2$ and pyrimidine cases has been found to be relatively small, especially when $\mathcal H$ coupling is applied.
%
%
The \acl{1CA} (\ac{1CA}) seems to be not applicable in all cases for the present method.
Although it is very attractive from the computational viewpoint as it substantially reduces the numerical effort. 
For example, the CH$_4$ \ac{1CA} results do not agree well to experiment, while for NO$_2$ they closely resemble the $r^{-1}$ ones.
However, for pyrimidine only the \ac{1CA} and \ac{2CA}, including two-center contributions from only heavy atoms, were feasible.
Unfortunately, these approximations do not lead to good agreement with experiment in this case; we suppose that the deviations can be rather attributed to deficiencies of the quantum chemistry setup, exclusion of hydrogen contributions, or the \ac{SCI} model itself.

Finally, some of the deviations between calculations and experiment can be mitigated if the one-step model is applied.
This might be especially needed to describe spectral regions with notable participator decay character and cases with strong interference between different ionization pathways. 
Taking into account nuclear motion might also be highly advisable.
\begin{acknowledgments}
 Financial support from the Deutsche Forschungsgemeinschaft Grant No. BO 4915/1-1 is gratefully acknowledged.
\end{acknowledgments}

\bibliography{bibtex/molecular_PES_AES.bib}

\end{document}